\documentclass[floatfix,lengthcheck,showpacs,amssymb,amsmath,amsfonts,twocolumn,nofootinbib,prd]{revtex4-1}

\usepackage{lineno}
\usepackage[utf8]{inputenc}
\usepackage{color}
\usepackage{graphicx}
\usepackage{float}
\usepackage{subfigure}
\usepackage{amsmath}
\usepackage{acronym}
\usepackage{multirow}
\usepackage{tikz}
\usetikzlibrary{arrows,shapes,trees,decorations.pathreplacing}
\usepackage{import}
\usepackage{svn-multi}
\usepackage{lineno}
\usepackage{hyperref}
\usepackage{gensymb}
\hypersetup{
    colorlinks=true,
    linkcolor=blue,
    filecolor=magenta,      
    urlcolor=cyan,
}
\usepackage[toc, nonumberlist]{glossaries}
\glsdisablehyper
\usepackage{verbatim}
\usepackage[binary-units=true, alsoload=astro]{siunitx}
\usepackage{bm}
\usepackage{mathtools}
\usepackage{rotating}

\begin{document}

\newglossaryentry{sr}{name=SR, description={Special relativity}}
\newglossaryentry{gw}{name=GW, description={Gravitational wave}}
\newglossaryentry{gws}{name=GWs, description={Gravitational waves}}
\newglossaryentry{gr}{name=GR, description={General relativity}}
\newglossaryentry{pn}{name=PN, description={Post-Newtonian approximation}}
\newglossaryentry{ligo}{name=LIGO, description=Laser Interferometer Gravitational-Wave Observatory}
\newglossaryentry{aligo}{name=aLIGO, description=advanced Laser Interferometer Gravitational-Wave Observatory}
\newglossaryentry{utc}{name=UTC, description=Coordinated Universal Time}
\newglossaryentry{dgl}{name=DGL, description=Differenzialgleichung/en}
\newglossaryentry{mf}{name=MF, description=Manifold}
\newglossaryentry{tt}{name=TT, description=Transversal-traceless gauge}
\newglossaryentry{obda}{name=o.B.d.A. , description=ohne Beschränkung der Allgemeinheit}
\newglossaryentry{onb}{name=ONB, description=Orthonormalbasis}
\newglossaryentry{bns}{name=BNS, description=Binary neutron star}
\newglossaryentry{bbh}{name=BBH, description=Binary black hole}
\newglossaryentry{isco}{name=ISCO, description=Innermost stable circular orbit}
\newglossaryentry{virgo}{name=VIRGO, description=Earthbound gravitational wave detector in Italy}
\newglossaryentry{pdf}{name=PDF, description=Probability density function}
\newglossaryentry{skp}{name=SKP, description=Skalarprodukt}
\newglossaryentry{gpu}{name=GPU, description=Graphics processing unit}
\newglossaryentry{pm}{name=PM, description=Post-Minkowskian approximation}
\newglossaryentry{eob}{name=EOB, description=Effective one body problem}
\newglossaryentry{nn}{name=NN, description=Neural network}
\newglossaryentry{nns}{name=NNs, description=Neural networks}
\newglossaryentry{cnn}{name=CNN, description=Convolutional neural network}
\newglossaryentry{cnns}{name=CNNs, description=Convolutional neural networks}
\newglossaryentry{ffn}{name=FFN, description=Feed forward (neural) network}
\newglossaryentry{rnn}{name=RNN, description=Recurrent neural network}
\newglossaryentry{rnns}{name=RNNs, description=Recurrent neural networks}
\newglossaryentry{snr}{name=SNR, description=Signal to noise ratio}
\newglossaryentry{mse}{name=MSE, description=Mean squared error}
\newglossaryentry{ilsvrc}{name=ILSVRC, description=ImageNet Large Scale Visual Recognition Challenge}
\newglossaryentry{tcn}{name=TCN, description=Temporal convolutional network}
\newglossaryentry{sgd}{name=SGD, description=Stochastic gradient descent}
\newglossaryentry{psd}{name=PSD, description=Power spectral density}
\newglossaryentry{tcin}{name=TCIN, description=TCN-collect-inception network}
\newglossaryentry{em}{name=EM, description=Electromagnetic}
\newglossaryentry{asd}{name=ASD, description=Amplitude spectral density}
\newglossaryentry{relu}{name=ReLU, description=Rectified linear unit}
\newglossaryentry{psnr}{name=pSNR, description=Peak signal-to-noise ratio}
\newglossaryentry{kagra}{name=KAGRA, description=Kamioka Gravitational Wave Detector}
\newglossaryentry{ns}{name=NS, description=Neutron star}
\newglossaryentry{roc}{name=ROC, description=Receiver operating characteristic}
\newglossaryentry{cpu}{name=CPU, description=Central processing unit}
\newglossaryentry{far}{name=FAR, description=false-alarm rate}

\title[]{Detection of gravitational-wave signals from binary neutron star mergers using machine learning}

\author{Marlin B. Sch{\"a}fer$^{1,2}$, Frank Ohme$^{1,2}$, Alexander H. Nitz$^{1,2}$
        }

\address{$^1$ Max-Planck-Institut f{\"u}r Gravitationsphysik,
         Albert-Einstein-Institut, D-30167 Hannover, Germany}
\address{$^2$Leibniz Universit{\"a}t Hannover, D-30167 Hannover, Germany}

\begin{abstract}
As two neutron stars merge, they emit gravitational waves that can potentially be detected by earth bound detectors. Matched-filtering based algorithms have traditionally been used to extract quiet signals embedded in noise. We introduce a novel neural-network based machine learning algorithm that uses time series strain data from gravitational-wave detectors to detect signals from non-spinning binary neutron star mergers. For the Advanced LIGO design sensitivity, our network has an average sensitive
distance of \SI{130}{\mega\parsec} at a false-alarm rate of 10 per month. Compared to other state-of-the-art machine learning algorithms, we find an improvement by a factor of 4 in sensitivity to signals with signal-to-noise ratio between 8 and 15. However, this approach is not yet competitive with traditional matched-filtering based methods. A conservative estimate indicates that our algorithm introduces on average \SI{10.2}{\second} of latency between signal arrival and generating an alert. We give an exact description of our testing procedure, which can not only be applied to machine learning based algorithms but all other search algorithms as well. We thereby improve the ability to compare machine learning and classical searches.
\end{abstract}

\maketitle

\section{Introduction}

The first direct detection of a gravitational-wave (\gls{gw}) signal on September the 14th 2015 \cite{gw150914} marked the dawn of gravitational wave-astronomy. During the first two observing runs, the \gls{ligo} and \gls{virgo} scientific collaboration found 11 \gls{gw}s \cite{catalog} from coalescing compact binary systems. Two independent reanalyses of the data have discovered a further set of events, 3 of which are found to be of astronomical origin with probability $p_\text{astro}>0.5$ by both studies \cite{ias, 2ogc, Nitz:2018imz}. The third observing run has identified tens of new \gls{gw} candidate events \cite{o3_alerts} and so far reported four new \gls{gw} detections \cite{gw190412, gw190425, gw190814, gw190521}. With detector sensitivity improving further for future observing runs and \gls{kagra} \cite{kagra, kagra2} joining the detector network, the rate of detections is expected to grow \cite{sensitivity_improvments_next_runs}.

The most sensitive low-latency searches are tailored specifically to signals from coalescing compact binaries and use a fixed number of pre-calculated templates \cite{ligo_pipelines}. 

Each template is a unique combination of a waveform model and source parameters. These searches work by calculating an inner product between the data and every template to produce a signal-to-noise ratio (\gls{snr}) time series. This process is known as matched filtering and is mathematically proven to be optimal for finding signals submerged in stationary, Gaussian noise \cite{gwv1}. 

If the SNR of a candidate exceeds a pre-selected threshold and the candidate is not excluded due to other factors, such as poor data quality or an implausible time of arrival difference between two different detectors, the low-latency search algorithms return a candidate event \cite{gstlal_pipeline, mbta_pipeline, pycbc_live, spiir_pipeline}.

The computational cost of a matched-filter search scales linearly with the number of templates used. This number will grow with the improving detector sensitivity at low frequencies \cite{template_number} of planned updates \cite{sensitivity_improvments_next_runs}. If currently neglected effects such as precession \cite{precessionSearch, firstPrecessionPhenom, latestPrecessionPhenom, firstPrecessionEOB},  higher order modes \cite{hmNeglected, hmModel, precessionHmPhenom, firstHmEOB} or eccentricity \cite{eccentricSearch} are taken into account, even more templates would be required. More computationally efficient algorithms would enable searching for sources which cannot currently be targeted due to a fixed computational budget.

When detecting \gls{gw}s from compact binary systems that contain at least one neutron star, the latency of the detection pipeline is critical, as these systems may produce electromagnetic (\gls{em}) signals. To detect these \gls{em} counterparts and maximize observation time, observatories need to be notified of possible events quickly. The number of false alarms on the contrary should be minimized as telescope time is expensive. Current low-latency searches introduce a latency of $\mathcal{O}\left(10\right)$ seconds and operate at a false-alarm rate (\gls{far}) of 1 per 2 months \cite{ligo_pipelines, gstlal_pipeline, mbta_pipeline, pycbc_live, spiir_pipeline}. Any new search needs to meet or exceed these standards to be considered for production use.

Neural network (\gls{nn}) based machine learning algorithms are an interesting alternative to traditional search algorithms, as they have shown great improvements in many tasks such as image recognition \cite{ILSVRC15}, sound generation \cite{wavenet} or certain board and computer games \cite{alpha_go, open_ai_five}. \gls{nn}s have also already found some application in the context of \gls{gw} data analysis~\cite{gravityspy, paper_christoph, dnn_denoising, Cuoco:2020ogp, Green:2020hst, Gabbard:2019rde, Marulanda:2020nww,Chan:2019fuz, Iess:2020yqj}. A few notable examples are  the classification of non-gaussian noise transients \cite{gravityspy}, the search for continuous \gls{gw}s \cite{paper_christoph} and denoising of detector data to recover injected signals \cite{dnn_denoising}. One key advantage of \gls{nn}s is their computational efficiency once trained. Most of the computational cost is shifted to the training stage, resulting in very quick evaluation. The application of \gls{nn}s to \gls{gw} searches might therefore offer a way to reduce computational cost of low-latency searches.

The authors of \cite{original_deep_filtering, hunter} were the first to directly apply deep \gls{nn}s to time series strain data to detect \gls{gw}s from binary black hole (\gls{bbh}) mergers. They tested the sensitivity of these searches at estimated \gls{far}s $\mathcal{O}\left(10^3\right)$ per month\footnote{We estimate this \gls{far} by multiplying the false-alarm probabilities given in \cite{original_deep_filtering, hunter} by the respective number of samples times the duration by which the position of the peak amplitude is varied within the training data-samples.}. Both analyses are able to closely reproduce the performance of a matched filter search at these \gls{far}s at a fraction of the computational cost. The \gls{nn}s excel at high \gls{far}s and low \gls{snr}. Both networks detected all signals with \gls{snr} larger 10 at estimated \gls{far}s of $\mathcal{O}\left(10^4\right)$ per month. These results are a promising first step but the algorithms would need to be tested at the required \gls{far}s of 1 per 2 months on real detector data to demonstrate an improvement over established methods.

Starting from their network, we reshaped the architecture significantly to optimize it to detect signals from binary neutron star (\gls{bns}) mergers. 
Our network-based search estimates the \gls{snr} and a quantity we call p-score for the given input data. The p-score is a measure for how likely the data is to contain a \gls{gw} signal. It is explicitly not a probability. The network is trained on simulated data of non-spinning binary neutron star systems with masses between $1.2$ and $1.6$ solar masses, isotropically distributed over the sky. 
All noise is stationary and Gaussian and as such does not contain any transients or other contaminations that are present in real detector data \cite{noise_transients_gw150914, o2_detchar, blip_glitches}. The previous works \cite{original_deep_filtering, hunter} have used data from a single detector. To improve the performance of our search, we expand the algorithm to work with data from two detectors. 
Using multiple detectors may also enable real-time estimates of the sky-position in the future.

Detecting \gls{bns} signals using a \gls{nn} is inherently more difficult than finding a \gls{bbh} signal, as (1) the \gls{gw} of a \gls{bns} reaches higher frequencies and (2) spends more time in the sensitive bands of the detectors. Due to (1) the data needs to be sampled at a high rate. Combined with (2) this leads to a massive increase of data that needs to be analyzed. As \gls{nn}s tend to be difficult to optimize when the input data has many samples, it is not feasible to na\"ively use the full time series sampled at a single rate as input. To solve this problem, we sample different parts of the signal at different rates. Frequencies emitted during the early inspiral are low and evolve slowly (see \autoref{fig:multirate}). High sample rates are only necessary during the final few cycles, where frequencies are high and grow rapidly.

The \gls{far}s probed by \cite{original_deep_filtering, hunter} are orders of magnitude larger than what is required for low-latency pipelines. Additionally these \gls{far}s were estimated on a discrete set of samples which either contain a signal or consist of pure noise. The waveforms within these samples are always aligned in a similar way and no signal is contained only partially in the analyzed segment. As the authors of \cite{cnn_magiacal_bullet} point out, \gls{far}s estimated on a discrete set of samples may for these reasons not be representative of a realistic search which has to work with a continuous stream of data.

We propose a standardized way of evaluating \gls{nn} \gls{far}s and sensitivities. To calculate these metrics, we generate a long stretch of continuous time series data which contains many injected \gls{gw}s that are roughly separated by the average duration of a \gls{bns} signal. Our network is applied to this data and points of interest are clustered into events. All results we provide are derived from analysis of $\approx 101$ days of simulated continuous data. We test the network down to \gls{far}s of $0.6$ per month and find sensitive distances of \SI{130}{\mega\parsec} down to \gls{far}s of $10$ per month.

We compare our search to the currently in-use low-latency detection pipeline PyCBC Live \cite{pycbc_live} and the results given by the authors of \cite{bns_network}, who were the first to classify \gls{bns} signals with a machine learning algorithm. We find an improvement in sensitivity of close to $400$\% for \gls{bns} signals with \gls{snr} in the range of provided training examples ($8\leq\text{\gls{snr}}\leq 15$) over the previous state-of-the-art machine learning algorithm. This makes our algorithm the best machine learning algorithm for detecting \gls{bns} signals at low \gls{snr}s. We are, however, not yet able to match the performance of template based searches. To do so we need to either increase the sensitive radius of our search at the lowest \gls{far}s by a factor of $6$ or double the sensitive radius while lowering the \gls{far} by an order of magnitude.

The trained network is public and can be found in the associated data release~\cite{bns-ml-release}. At the same location we also provide example code of how to apply it to long stretches of data and a function that generates injections as they are used to derive \gls{far}s and sensitivities in this work.

The contents of this paper are structured as follows: Section \ref{sec:testing_methodology} describes how search algorithms should be evaluated. It gives the general concepts in the first part and details on how to apply these concepts to \gls{nn}s in the second part. Section \ref{sec:training_validation_data} explains the multi-rate sampling and describes the data used for both training and validation of the network. The following section \ref{sec:network} gives an overview of the architecture and how this architecture is trained and tested. We present our results in section \ref{sec:results} of which we draw conclusions in section \ref{sec:conclusion}.

\section{False-alarm rate and Sensitivity  of gravitational-wave search algorithms}\label{sec:testing_methodology}

There are two important metrics that have been used to evaluate gravitational-wave searches in the past. These two are the \gls{far} of the search and the corresponding sensitivity \cite{pycbc_search}. In principle, these metrics can directly be applied to \gls{gw} searches that utilize \gls{nn}s. As pointed out by the authors of \cite{cnn_magiacal_bullet}, in practice the discrete nature of the data that is used to train these networks has lead to some divergence between the terminology used for \gls{nn} and traditional search algorithms.

\subsection{Calculation for general search algorithms}\label{sec:evaluateion_general}
The main goal of a search algorithm is to detect \gls{gw} signals in real data,
where the input is a nearly continuous strain time series. A search therefore must produce a list of times of candidate events and rank them by a \textit{ranking statistic} $\mathcal{R}$. The ranking statistic is a number which signifies how likely the data is to contain a signal. To evaluate the performance of an algorithm, it is applied to mock data containing known \textit{injections}, i.e. additive \gls{gw} signals with known parameters. The events generated from this data are compared to the list of injections and used to determine which injected signals were found, which missed and which events are false alarms. 

Any event that is reported by the search needs to be assigned a \gls{far} to express the confidence in its detection. For a given value $\mathcal{R}$, the \gls{far} is the number of false alarms with a ranking statistic of at least $\mathcal{R}$ per unit time. To estimate it on mock data, the number of false detections exceeding a ranking statistic $\mathcal{R}$ is divided by the duration of the analyzed data.

The ability of the search to recover signals is quantified by the sensitivity which is a function of the \gls{far} lower bound. It is often given in terms of the fraction of recovered injections. This fraction, however, strongly depends on the parameter distribution of the injected signals, as the amplitude of the signal in the detector depends on the orientation and location of the source. Thus, the fraction can be diminished by injecting sources at larger distances or unfavorable orientations. A more astrophysically motivated measure of sensitivity is the sensitive volume of the search algorithm. It is an estimate of the volume around the detector from which \gls{gw} sources will be detectable. This volume may be calculated through
\begin{equation}\label{def:sensitive_volume}
V\left(\mathcal{F}\right)=\int \mathop{}\!\mathrm{d}\mathbf{x}\mathop{}\!\mathrm{d}\mathbf{\Lambda}\  \epsilon\left(\mathcal{F};\mathbf{x},\mathbf{\Lambda}\right)\phi\left(\mathbf{x},\mathbf{\Lambda}\right),
\end{equation}
where $\epsilon\left(\mathcal{F};\mathbf{x},\mathbf{\Lambda}\right)$ is the efficiency of the search pipeline for signals with \gls{far} $\mathcal{F}$, spatial co-ordinates $\mathbf{x}$ and injection parameters $\mathbf{\Lambda}$. The function $\phi\left(\mathbf{x},\mathbf{\Lambda}\right)$ is a probability density function which describes the astrophysical distribution of signals \cite{pycbc_search}. When the distribution of injections matches the expected astrophysical distribution (i.e. uniform in volume, isotropic in sky location, etc.), equation \eqref{def:sensitive_volume} can be estimated by

\begin{equation}\label{def:sensitive_volume_estimate}
V\left(\mathcal{F}\right)\approx V\left(d_\text{max}\right)\frac{\text{\# found injections}(\mathcal{F})}{\text{\# total injections}},
\end{equation}
where $d_\text{max}$ is the maximal distance of injected signals, $V\left(d_\text{max}\right)$ is the volume of a sphere with radius $d_\text{max}$ and the function $\text{\# found injections}(\mathcal{F})$ counts the number of detected injections with a \gls{far} $\leq\mathcal{F}$. 
 
We use the function \verb|volume_montecarlo| of the PyCBC software library \cite{pycbc} to carry out this estimation.

Current searches notify astronomers of a \gls{gw} event when the event is assigned a \gls{far} of at most 1 per 2 months \cite{pycbc_live}. Any new search should hence be tested at least down to these \gls{far}s. To resolve \gls{far}s of that scale at least 2 months of mock data are required.

For our tests we generated 100 files each containing roughly 1 day of continuous data. Each file contains independently drawn data. For easier multiprocessing, each file is internally split into 22 chunks of duration 4096 seconds. We start by generating a list of injection times, requiring that two injections are separated by 180 to 220 seconds. The exact separation time is chosen uniformly from this interval. To avoid waveforms that are not completely within one chunk we discard any injections that are within the first or final 256 seconds of each chunk. For every injection time, we generate a waveform using the inspiral-only waveform model \verb|TaylorF2|~\cite{taylorf2_1, taylorf2_2, taylorf2_3} with a lower frequency cutoff of 25 Hz. Its parameters are drawn from the distribution specified in \autoref{tab:parameter_distribution}. Finally, the waveform is projected into the frame of the LIGO-Hanford and LIGO-Livingston detectors and added into simulated Gaussian noise such that the peak amplitude is positioned at the injection time. All noise is generated from the analytic estimate of the power spectral density (\gls{psd}) of the \gls{aligo} final design sensitivity as provided by the software library LALSuite \cite{lalsuite}.
\begin{table}
\centering
\begin{tabular}{c|c}
    parameter & uniform distribution\\
    \hline\\
    component masses & \SI[parse-numbers=false]{m_1, m_2\in\left(1.2,1.6\right)}{M_\odot}\\
    spins & 0\\
    coalescence phase & $\Phi_0\in\left(0, 2\pi\right)$\\
    polarization & $\Psi\in\left(0, 2\pi\right)$\\
    inclination & $\cos{\iota}\in\left(-1, 1\right)$\\
    declination & $\sin{\theta}\in\left(-1, 1\right)$\\
    right ascension & $\varphi\in\left(-\pi, \pi\right)$\\
    distance & \SI[parse-numbers=false]{d^2\in\left(0^2, 400^2\right)}{{\mega\parsec}^2}
\end{tabular}
\caption{The astrophysically motivated distribution of parameters used to generate injections. These are used to estimate the \gls{far} and sensitivity of the search algorithm specified in this paper.}\label{tab:parameter_distribution}
\end{table}

\medskip

\subsection{Calculation for neural network searches}\label{sec:evaluation_neural_networks}
A \gls{nn} is a general purpose function approximator that finds a fit for a set of example input-output pairs. This fitting process is called training and the example data used is called the training set. Once trained, the network can be applied to data that was not covered by the training set and will evaluate the fit function at this new point. It does so assuming that the unseen data-samples originate from the same underlying process as the training data-samples. The given output for any unseen data-sample is thus an interpolation or extrapolation of the example outputs from the training set.

To generate \gls{far}s and sensitivities, previous works \cite{hunter,original_deep_filtering} generated a second set of noise- and signal-samples with the same duration and sample rate used in the training set. They then applied the network to this second set of data-samples and determined \gls{far}s by counting how many noise-samples were classified as signals and sensitivities by counting how many signal samples were classified as such.

There are two main problems with using a discrete set of data-samples to determine \gls{far}s and sensitivities. The first stems from the structure of the data-samples themselves. To make the training process more efficient, it is necessary to position the peak amplitude of the \gls{gw} signal within a narrow band of the data-sample. When applied to real data, this property can not be ensured, and the assumption of the data being similar to the training set is not well approximated. Hence, if a \gls{far} or sensitivity is calculated on a set where the alignment is guaranteed, it will not necessarily be representative of the performance on realistic data. The second problem is the required fixed duration of input data-samples. Usually, a search algorithm is applied to long stretches of time series data to find potential signals. To evaluate data of greater duration than the accepted input size of the network, it is applied multiple times via a sliding window. At each position the output will give an estimate if a signal is present. If this is the case, it is initially not clear what a true positive is, as the network may predict the presence of a signal for multiple consecutive positions, the input window may only partially contain a signal or the network may jump between predicting the presence and absence of a signal for multiple subsequent positions \cite{cnn_magiacal_bullet}.

To generate representative \gls{far}s and sensitivities, we propose to use mock data of much greater duration than the input duration of the network. The network is then applied to this data by sliding it across. The step size should at most be half the size of the interval where peak amplitudes of the waveforms occur in the training set. This step size ensures that any waveform is positioned correctly for at least one position of the network. 

If the output of the network is not binary but continuous it can be interpreted as a ranking statistic. In this case a threshold can be applied to find positions where the network predicts to have found a signal. 
Candidate events are identified in the resulting time series by applying a threshold and clustering.

Each event is assigned a specific time and ranking statistic. The resulting list of events is compared to the list of known injections as described in \ref{sec:evaluateion_general} to calculate \gls{far}s and sensitivities. The specifics of our analysis and clustering algorithm are described in \ref{sec:test_bns}.

As every event needs to be assigned a ranking statistic, we can calculate the metrics by only using events that exceed a given threshold. Doing so for many different values allows us to obtain the \gls{far} as a function of the ranking statistic threshold and subsequently also the sensitivity as a function of the \gls{far}.

We found that testing our \gls{nn} on long stretches of data and applying clustering to obtain a list of events increased the \gls{far} over \gls{far}s measured on a set of discrete samples at the same detection threshold by at least a factor of 2. To give comparable statistics, we therefore strongly recommend to test networks in the way described above.

\section{Data processing}\label{sec:training_validation_data}

To train and evaluate the network, three disjoint data sets with known contents are required. One of the three sets is the training set and used to optimize the parameters of the network. The network is usually trained on the training set multiple times, where each complete pass is called an epoch. After multiple epochs, the network may start to learn specifics of the provided data-samples rather than the general structure. This behavior is called overfitting and can be detected by monitoring the performance of the network on a validation set. Different epochs are rated by their performance on this second set.  Ranking the different training stages of the network in this way introduces a selection bias and optimizes the network on the validation set. To give unbiased results a testing set is required. This one should represent the data that the network will be applied to the closest and should optimally be generated independently from the training and validation set. To keep results as unbiased as possible, the testing set should ideally only be analyzed once. This section describes how an individual data-sample needs to be formatted for our network and how the training and validation set are generated. Details on the testing set are described in \autoref{sec:evaluateion_general}.

\subsection{Input data preparation}\label{sec:data_processing}

Previous works \cite{original_deep_filtering, hunter, huerta_parameter_estimation} have already successfully classified whitened time series data for \gls{bbh} signals with a simple convolutional neural network. As input they used 1 second of data sampled at \SI{8192}{\hertz}. For \gls{bbh} signals this is a sensible choice for the duration of analyzed data, as these signals sweep through the sensitive frequency range of the detectors in $\mathcal{O}(1)$ seconds, and the chosen sample rate is sufficient to resolve them. Signals from binary neutron star mergers, on the other hand, spend $\mathcal{O}(100)$ seconds in the sensitive band of the detectors. Using the usual signal duration as input to the network would lead to a hundredfold increase of data-points over the \gls{bbh} case. Training a \gls{nn} with this many input samples is infeasible due to memory constraints and optimization problems.

To reduce the number of input samples, the authors of \cite{bns_network} use only the final 10 seconds of each signal as input. This enables them to differentiate not only between noise and \gls{bns} signal, but also distinguish \gls{gw}s from \gls{bbh} mergers. They test an \textit{architecture}, i.e. a network structure, similar to those of \cite{original_deep_filtering, hunter} and are able to closely reproduce their results for \gls{bbh} data. Their sensitivity to \gls{bns} signals looks very promising, but the search has yet to be tested at realistic \gls{far}s. The short duration of the input in comparison to the duration a \gls{bns} signal spends inside the sensitive band of the detectors reduces the \gls{snr} contained in the data by about $25\%$ and thus limits the sensitivity of the search.

\begin{figure}
    \centering
    \includegraphics[width=0.45\textwidth]{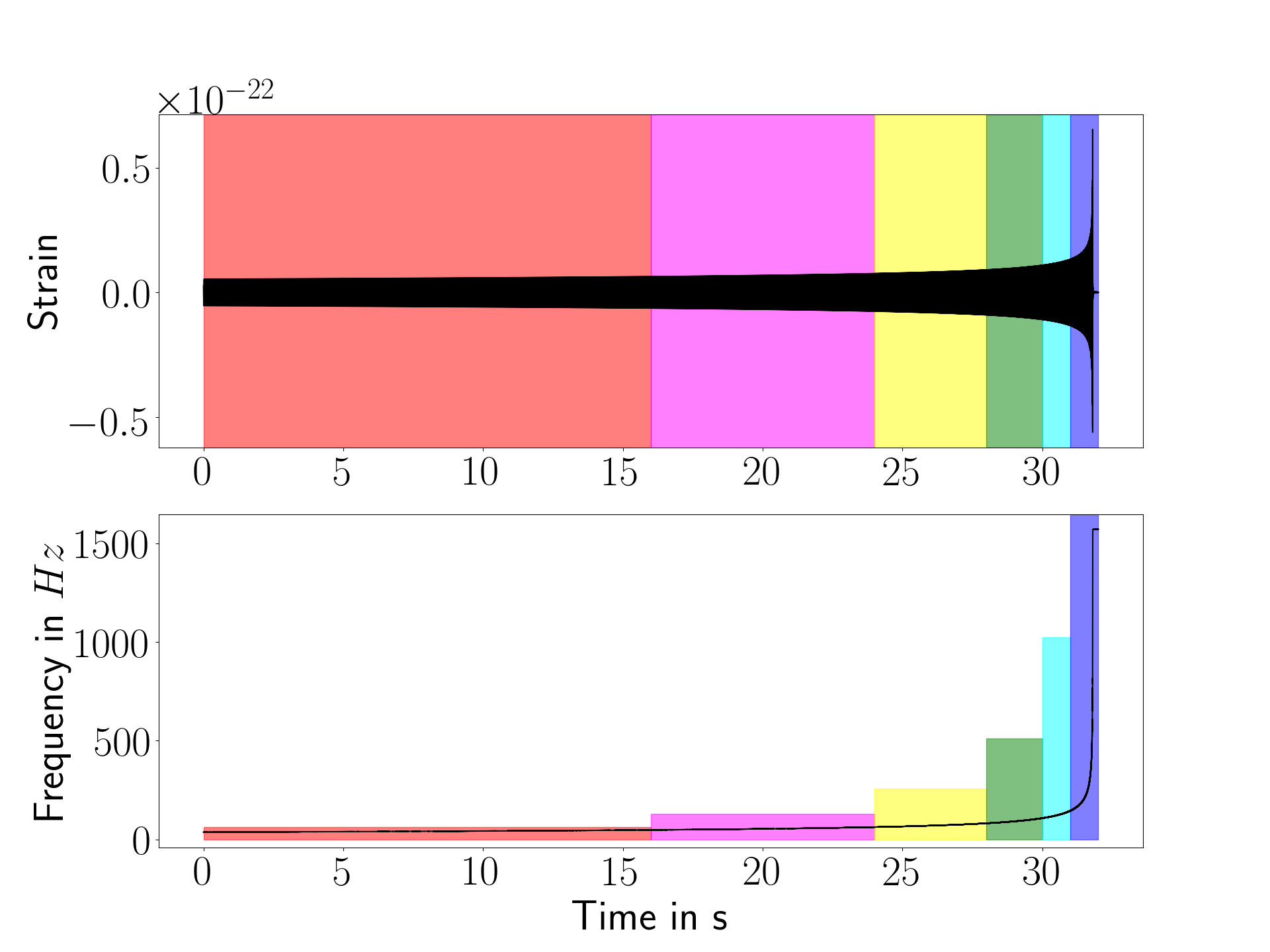}
    \caption{The top panel shows the strain evolution of an example \gls{gw} from a \gls{bns} merger in black. The bottom panel shows the corresponding frequency evolution in black. The colored boxes represent parts of the signal which we sample at different rates. The height of these boxes in the bottom panel represents the Nyquist-frequency of the sample rate which is used for each part. To fully resolve the signal, the black curve must stay inside the colored boxes of the bottom panel at all times.}
    \label{fig:multirate}
\end{figure}

To retain as much \gls{snr} in the data as possible while at the same time reducing the input size to the network, we sample 32 seconds of data containing a potential signal at different rates. During the early inspiral, frequencies are low and the frequency evolution is slow. This allows us to sample a long stretch in the beginning at a low rate. The first 16 seconds are sampled at \SI{128}{\hertz}, the following 8 seconds are sampled at \SI{256}{\hertz} and so on. The pattern continues until the final second, which is sampled at \SI{4096}{\hertz} but split into two parts of equal length. We ensure that no two sections overlap to reduce redundant information. This method of re-sampling generates 7 parts containing 2048 samples each and ensures that for every part of the signal a sufficient sample rate is used. The number of samples is reduced by a factor of 9 (See \autoref{fig:multirate}) and about $98\%$ of the \gls{snr} is retained.\footnote{Notice that this way of sampling the data differs from our previous work \cite{master} in that we dropped the lowest sample rate. We found that for some signals a sample rate of \SI{64}{\hertz} for the first 32 seconds of a 64 second signal was not sufficient to resolve the highest frequencies during that stage of the binary evolution and introduced unwanted artifacts. We also sampled the \gls{psd} used for whitening the data too coarsely in our previous work \cite{master} and signals were more difficult to find as a result.} 

Rather than re-sampling the data directly, we first \textit{whiten} it using the analytic model \verb|aLIGOZeroDetHighPower| for the aLIGO design sensitivity \gls{psd} as provided by the software library LALSuite \cite{lalsuite}. Whitening of data is a procedure where every frequency bin is reweighted by the average power of the background noise in this bin. It ensures that power in any frequency bin in excess of unity is an indication for the presence of a signal. For computational efficiency during training, noise and signal samples are whitened individually. Since the whitening procedure is a linear operation, whitening the sum is equivalent to whitening both parts individually. Both parts are combined at runtime on the first layer of the \gls{nn}. The reason to store them separately is an increase in the effective number of samples which can be achieved by mixing and matching different signals and noise samples. It also helps to improve the efficiency of training by using the same signal template submerged in different realizations of noise.

When evaluating real samples, we cannot trivially separate the signal from the background and thus cannot whiten each part individually. Instead we whiten the total signal by the same analytic model of the \gls{psd} used for the training and validation data. The whitened data is re-sampled and used as the signal input. As the signal input already contains the total signal including noise and the network only sees the sum of both inputs, the noise input is set to zero.

\subsection{Generating training and validation set}\label{sec:data_generation}
All signals for the training and validation set are generated using the inspiral-only waveform model \verb|TaylorF2| from the software library LALSuite \cite{lalsuite} with all parameters but the distance drawn from the distribution given in \autoref{tab:parameter_distribution}. The luminosity distance $d$ is set indirectly by uniformly drawing a target network \gls{snr} from the interval $\left[8,15\right]$. The waveform is first computed at a fiducial distance of \SI{1}{\mega\parsec} with a low-frequency cutoff of \SI{20}{\hertz}. Then the waveform is projected onto the two detectors Hanford and Livingston \cite{signal_distribution} and cropped to a length of 96 seconds. During this step, we shift the waveform such that the peak amplitude occurs within the final $4.25$ to $4.75$ seconds\footnote{Altering the time of the peak amplitude of the waveform during training allows the network to be less sensitive to the exact position of the waveform within the input window. This enables us to slide the network across long stretches of input data with a larger step size. For this study, we chose to use an interval of \SI{0.5}{\second}. It may be possible to optimize this choice, but we have not done so here.} of the 96 seconds data-segment. The exact position within this interval is drawn uniformly. Next, we calculate the network \gls{snr} by taking the root of the sum of the squares of the inner product of the waveforms with themselves \cite{signal_distribution}, weighing each frequency by the analytic \gls{psd} \verb|aLIGOZeroDetHighPower| of \cite{lalsuite}. The waveforms are finally scaled by multiplying with the target network \gls{snr} and dividing by the network \gls{snr} at distance \SI{1}{\mega\parsec}. Afterwards the data is whitened, the initial and final $4$ seconds are discarded to avoid filter wrap-around errors, and the last $32$ seconds of the remaining segment are re-sampled as described in \ref{sec:data_processing}. Noise samples are simulated from the analytic \gls{psd} used above, whitened and re-sampled. As such all noise is stationary and Gaussian.

The training set contains 25,000 different \gls{gw} signals and 80,000 instances of noise. When training the network we pre-select 800,000 unique combinations of signal and noise at random and shuffle them with all 80,000 noise samples to obtain 880,000 training samples with a $10:1$ split between signals and noise. To compensate for this inequality we apply \textit{sample weights} of $1/10$ to all signal samples during training. Sample weights modify the loss by reweighting contributions from the according sample.

The validation set contains 1,500 different \gls{gw} signals and 8,000 instances of noise. We again generate 24,000 unique combinations of $\text{signal}+\text{noise}$ and shuffle them with the 8,000 noise samples. This results in a validation set that contains 32,000 samples with a $3:1$ split for $\text{signal}:\text{noise}$.

\section{The search algorithm\label{sec:network}}
\subsection{Neural network architecture}\label{sec:architecture}
When working with neural networks, the details of the implementation of the machine learning algorithm are mostly defined by the architecture of the network. There is no known optimal procedure for designing a network that works well for a given problem.

The architecture presented in this paper is highly optimized for the problem of detecting \gls{bns} signals and relies on the input data format described in \ref{sec:data_processing}. Some of the main improvements over a standard convolutional architecture will be more general and may be of use for different data formats and similar problems.

We started our research by adjusting the architecture given in \cite{original_deep_filtering, hunter} for data sampled at multiple rates, by using one channel for every sample rate and detector combination. In convolutional networks, channels represent different features of the data and are correlated by the convolutional filters. With this as a starting point, we made iterative improvements. The three changes that had the greatest positive effect were the replacement of convolutional layers by inception modules \cite{inception_module}, the use of a temporal convolutional network (\gls{tcn}) as a signal amplifier \cite{tcn_paper, tcn_idea, dnn_denoising} and using different stacks of inception modules for each sample rate. A detailed description of the evolution of the network can be found in \cite{master}. The architecture presented here differs from an earlier iteration presented in \cite{master} only by removing the lowest sample rate as input and adjusting the structure accordingly.

\begin{sidewaysfigure*}
    \centering
    \vspace*{9.5cm}\includegraphics[width=0.9\textwidth]{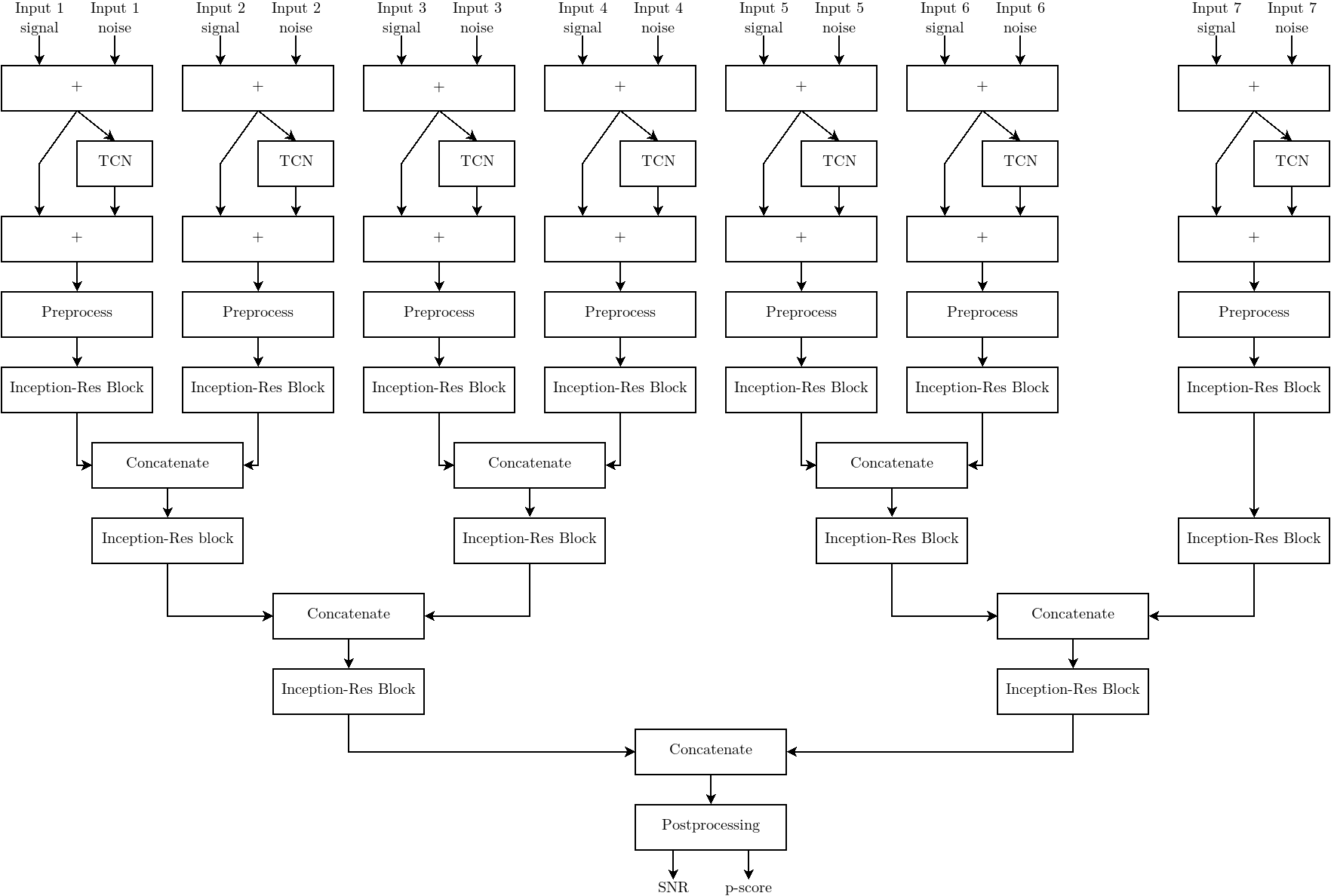}
    \caption{A high level overview of the architecture presented in this work. Details on every block can be found in \cite{master}. The network takes signal and noise inputs $1$ to $7$, where each number corresponds to a different part of the re-sampled raw data described in \ref{sec:data_processing}. It outputs an estimate of the \gls{snr} contained in the input and a p-score, which rates how likely the data is to contain a \gls{bns} signal. All auxiliary outputs that are only used for training are not shown. The network adds the noise and signal input for every re-sampled part individually and the remaining layers operate only on this sum. The output of this addition is amplified by a \gls{tcn} and processed by an inception network. Afterwards the outputs of two inception networks from adjacent sample rates are concatenated and further analyzed by another inception network. The parallel inception networks are concatenated until only a single one remains. A few final dense layers, which are summarized as the post-processing block, are applied to reduce the output shape to the desired dimensions of the \gls{snr} estimate and p-score. The pre-processing block is inspired by \cite{inception_module} and contains a small convolutional network.}
    \label{fig:architecture}
\end{sidewaysfigure*}

For computational efficiency, we provide the noise and signal time series not as a sum but as separate inputs to the network. They are combined on the first layer of each parallel stack of layers (see \autoref{fig:architecture}). This sum is passed to a \gls{tcn} which tries to recover the pure signal. The denoised data is added back onto the input of the \gls{tcn} to amplify potential signals. The amplified data is preprocessed by convolutional layers before 2 inception modules with very small kernel sizes are applied. Afterwards two adjacent stacks are concatenated and used as input to further inception modules. The process is repeated until only a single stack is left. This stack is reduced down to the desired output format by applying dense layers. A high-level overview of the architecture can be found in \autoref{fig:architecture}.

Inception modules are the main building block of the network. Their use was motivated by recent developments for image recognition tasks \cite{inception_module, ILSVRC15}.

The main advantage of inception modules over convolutional layers is the reduced number of parameters. In convolutional layers often many weights are close to zero after training. Ideally, sparse operations would be used in such cases. Sparse operations are, however, not computationally efficient. The idea of inception modules is to combine many small, dense filters in parallel to form an effective large, sparse filter. This approach allows for the use of efficient dense operations while reducing the number of trainable parameters at the same time \cite{inception_module}.

The final outputs of our network are one scalar for the \gls{snr} estimate and a tuple of length 2 estimating the p-score. The p-score is a measure for how confident the network is that the data contains a \gls{gw} and the content of the corresponding tuple is $(\text{p-score}, 1-\text{p-score})$ for technical reasons. For the training and validation set signal samples are labeled with a p-score of 1 and noise samples are labeled with a p-score of 0. The network output on the other hand is not binary but continuous. We thus  interpret both the \gls{snr} and the p-score output as ranking statistics.

Alongside the two outputs described above, the network is equipped with 13 further auxiliary outputs. The purpose of these outputs is to prevent vanishing gradients \cite{inception_module} or provide the intermediate layers with more information on their purpose. The auxiliary outputs thus improve the training efficiency of the network. Seven of the auxiliary outputs are the outputs of the \gls{tcn}s. They are trained using the pure signals as target. We found that the network is significantly more sensitive if it cannot decide how to use the parameters of the \gls{tcn}s freely but is forced to learn to recover the \gls{gw}. Five of the remaining six outputs are taken after all but the final concatenation layer. They receive the injected \gls{snr} as target. Since the output of the concatenation layers is not a scalar, we use a few pooling- and dimensional reduction layers \cite{dim_red_invention, master} to reduce the output shape. The final auxiliary output is taken after the first two inception modules of the lowest sample rate and treated in the same way as the auxiliary outputs after the concatenation layers. The network is trained as a complete unit and the loss takes into account all $15$ outputs.

The complexity of this architecture comes at the cost of memory size and speed. The model has $2.5$ million trainable parameters. The computational cost is a problem when optimizing the architecture as it is costly to compare two different designs. We therefore suspect that the details of this architecture can be further optimized.

\subsection{Training}
The network requires \SI{17}{\giga\byte} of memory to be trained with a mini-batch size of 32. We used a NVIDIA V-100 \gls{gpu} with \SI{32}{\giga\byte} of video memory for training. On this hardware each training epoch plus the subsequent validation step takes roughly 5 hours to complete. Due to time constraints and instabilities during training the network is trained for 24 epochs only. The instabilities are discussed below and manifest as a sudden drop of the sensitivity during training.

The total loss of the network is the weighted sum of the individual losses for each of the outputs discussed in \ref{sec:architecture}. All auxiliary outputs are assigned a mean squared error as the individual loss and given a weight of $0.1$. The same loss function is used for the \gls{snr} output, but it receives a weight of 1. Finally the p-score output uses the categorical cross-entropy and a weight of 0.5. The total loss is thus given by
\begin{align}
L\left(\bm{y}_\text{true}, \bm{y}_\text{pred}\right) & =\text{MSE}\left(\text{\gls{snr}}_\text{target}, \text{\gls{snr}}_\text{pred}\right)\nonumber\\
& +\frac{1}{2}\sigma\left(\text{p-score}_\text{target}, \text{p-score}_\text{pred}\right)\nonumber\\
& +\frac{1}{10}\sum_{i=1}^{13} \text{MSE}\left(y_{i,\text{target}}, y_{i, \text{pred}}\right),
\end{align}
where $\text{MSE}\left(x,y\right)\coloneqq{\left(x-y\right)}^2$ is the mean squared error, $\sigma\left(x,y\right)\coloneqq-\sum_{i=1}^2 x_i\log\left(y_i\right)$ is the categorical cross-entropy, a subscript ''target'' indicates the known target values, a subscript ''pred'' indicates the network output and the $y_i$ are the auxilliary outputs.
The different weights are used to inform the optimization algorithm on the importance of each individual output. The auxiliary outputs are only used during training and discarded during inference. Their value is unimportant as long as using them improves the performance of the \gls{snr} and p-score output. We use the default implementation of the ''Adam'' optimizer from the machine learning library Keras \cite{keras} to train the entire network using the total loss. Altering the initial learning rate in either direction reduced the sensitivity of the network.

During training we monitor an estimate of the sensitivity of our network. To do so, we calculate the true positive rate on the validation set, by choosing the maximum predicted \gls{snr} and p-score value of all noise samples from the validation set as a threshold. All signals that are estimated with a ranking statistic higher than these thresholds are counted as detected. The number of detected signals is then divided by the total number of signal samples to get a true positive rate. We rank the epochs based on this estimate of the sensitivity and thoroughly test the best one.

We found that the network experienced strong overfitting. While the training loss fell by 25\% from the initial to the last epoch, the validation loss doubled. If the loss and the sensitivity were strongly correlated it would be expected that the sensitivity drops with an increasing validation loss. We find the opposite and reach the highest true-positive rate of $16.5$\% on epoch 21. At this point the validation loss grew by 75\% over the initial epoch. The loss in use is therefore at best loosely correlated with the sensitivity. Designing a loss function that is better suited to the problem might improve the search further. The strong overfitting also indicates the possibility to simplify the architecture significantly without a strong loss of sensitivity or improving the performance of the current architecture by increasing the size of the training set significantly.

When training networks that predict if a \gls{gw} signal is present in some time series data, we found that after some number of epochs the sensitivity estimate drops to zero for both the \gls{snr} and the p-score output. Initially, it often recovers on the next epoch, but drops become more frequent. After some point the network does not seem to recover at all and the estimated sensitivity stays at zero. This behavior is caused by noise samples that are estimated with very high confidence to contain a \gls{gw}. These are sometimes appointed physically nonsensical \gls{snr} values. The number of these misclassified noise samples is low and thus the impact on the total loss is negligible. Furthermore, the values that are given for these noise samples grow over time, which is the reason why the drop occurs only after training for a while. In principle, these outliers may be vetoed by their \gls{snr} value at the cost of some sensitivity at low \gls{far}s. We disfavor this approach as it introduces artificial constraints on the search algorithm. It is currently unknown what causes the predicted values of the ranking statistics to grow or how the issue can be resolved. To avoid problems, we stop training before the sensitivity estimate stays at zero for many consecutive epochs.

Previous works \cite{original_deep_filtering,bns_network} have reported that a training strategy known as \textit{curriculum learning} improved the performance of their networks. During curriculum learning the network is trained on high \gls{snr} examples in the beginning, with training samples getting more difficult to differentiate from noise during later epochs. We tested this approach during early stages of our research and could not find a significant impact on the performance of our networks at the time. We therefore did not pursue the strategy during later stages of our project. However, we do suspect that curriculum learning might help to mediate the deficiencies of our network for loud signals.

\subsection{Testing on binary neutron star injections}\label{sec:test_bns}

To evaluate the sensitivity of the network, we use the test data described in \ref{sec:evaluateion_general}. It contains 8,794,112 seconds $\approx$ 101 days of data split into 100 files. With this data set, \gls{far}s down to $\approx 0.3$ false alarms per month can be resolved.

To analyze continuous stretches of data, we use a sliding window of duration 72 seconds with a step size of $0.25$ seconds. We chose the step size based on the training set in which the exact position of the peak amplitude was varied by $\pm 0.25$ seconds around a central position.

The content of every window is whitened by the analytic model \gls{psd} of the advanced \gls{ligo} detectors as provided by the software library LALSuite \cite{lalsuite}. To avoid filter-wraparound errors the initial and final 4 seconds are discarded. The final 32 seconds of the remaining data are re-sampled and formatted as described in \ref{sec:data_processing}.

To assign the correct times to each window, the alignment of the waveforms in the training set needs to be considered. The central position for the peak amplitude in the training set is set to $0.5$ seconds from the end. If the merger time is defined as the time of the peak amplitude of the waveform it will on average be positioned $31.5$ seconds from the first sample of the 32 second input window. Considering the $36$ seconds that are discarded at the beginning of each window, the first position of a \gls{gw} merger we are sensitive to is located at $67.5$ seconds from the start of each continuous segment. The reported sample times are therefore
\begin{equation}
    t(n) = t_\text{start} + 67.5\ \text{seconds} + (n-1)\cdot 0.25\ \text{seconds},
\end{equation}
where $t(n)$ is the time of the $n\text{-th}$ sample and $t_\text{start}$ is the starting time of the analyzed data segment.

By applying our network in this way, we obtain two time series. One estimates the \gls{snr} at each window position while the other gives a p-score at every step. We apply a fixed threshold of \gls{snr} $4$ and p-score $0.1$ to the respective time series. Every position that exceeds the corresponding threshold is marked. All marked positions are then clustered by assuming that two marked positions are generated by the same underlying process if they are within 1 second of each other. The clusters are expanded until there are no marked positions within 1 second of the boundaries of the cluster. Each cluster is an event and assigned the time and value of the maximum \gls{snr} or p-score respectively inside this cluster. An event is said to be a true positive if an injection was placed within $\pm 3$ seconds of the reported time. The times used for clustering and accepting a signal as true positive were empirically found to work well on a different data set and are arbitrary choices.

\section{Results}\label{sec:results}
\subsection{False-alarm rate and sensitivity}
\begin{figure}
    \centering
    \includegraphics[width=0.45\textwidth]{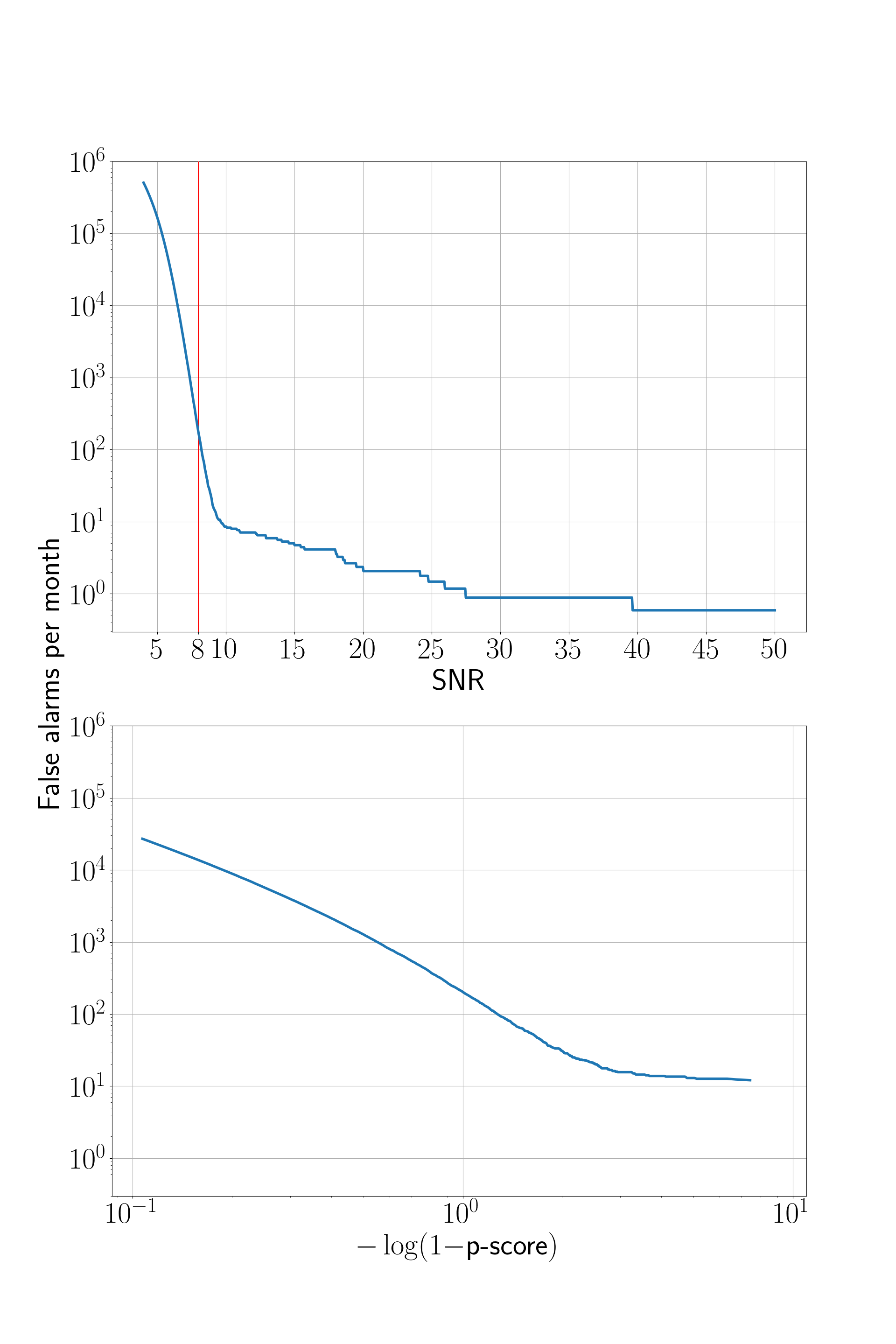}
    \caption{The estimated \gls{far} as a function of the threshold value used for either output. The top panel shows the \gls{far} of the \gls{snr} output. The red line in this plot points out the lowest \gls{snr} of training samples. In our previous work \cite{master} we found a change in gradient at this position. For the current search this change appears at a higher \gls{snr}. The bottom panel shows the \gls{far} of the p-score output. It is logarithmic and is scaled to give a sense for the distance to $\text{p-score}=1$.}
    \label{fig:false-alarm}
\end{figure}

\begin{figure}
    \centering
    \includegraphics[width=0.45\textwidth]{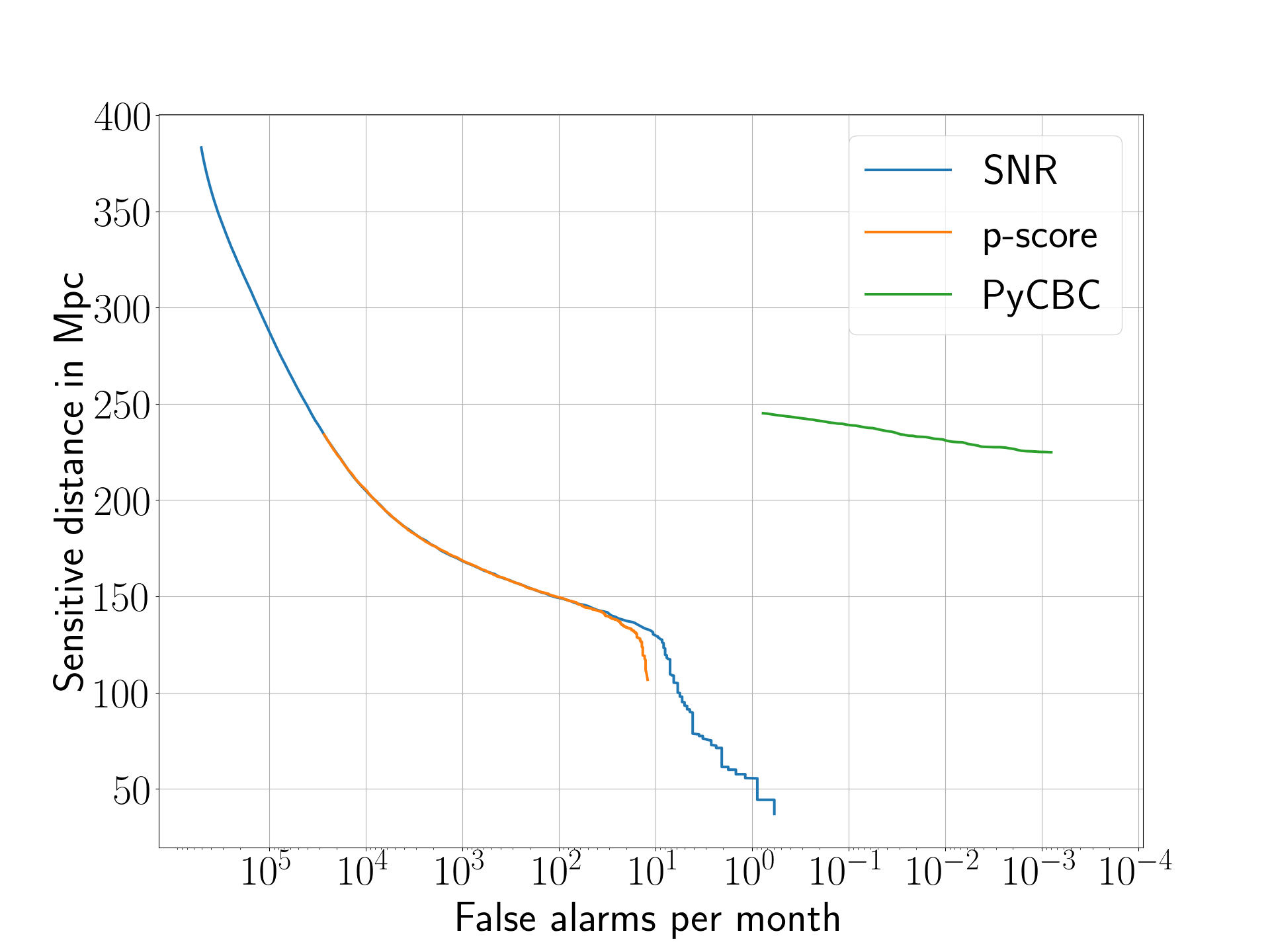}
    \caption{The sensitive distance as a function of the \gls{far}. The blue curve shows the sensitive distance when the \gls{snr} is used to classify events. The yellow curve shows the sensitive distance when the p-score is used. The green curve is generated from the data found in \cite{pycbc_sensitivity_plot} by counting all signals at a higher injection \gls{snr} than the corresponding \gls{far}. We are able to resolve a small overlap-region between the two different searches but find that the sensitivity of our search drops close to zero for \gls{far}s below 10 per month. At high \gls{far}s both outputs of our network perform equally well, for low \gls{far}s the \gls{snr} shows superior performance.}
    \label{fig:roc}
\end{figure}

The analysis of the \gls{bns} test data described in \ref{sec:test_bns} returns a list of events. Each event is assigned a ranking statistic. We obtain the \gls{far} as a function of the ranking statistics \gls{snr} and p-score (\autoref{fig:false-alarm}) by considering only those events that exceed the given threshold. Subsequently we can generate the sensitivity as a function of the \gls{far} (\autoref{fig:roc}). We choose a range of SNR 4 to 20 and p-score 0.1 to 1 to generate these plots.

We find that the \gls{snr} estimate is able to resolve \gls{far}s down to $0.6$ per month, whereas the p-score output is able to resolve \gls{far}s down to $12$ per month. Both curves drop steeply with the corresponding ranking statistic until they reach a \gls{far} $\mathcal{O}(10)$. At this point both curves level off significantly. Our previous work \cite{master} was able to resolve \gls{far}s down to $\approx 30$ per month and was tested on a set of roughly half the duration used in this paper. We also observed a change in gradient of the \gls{far} in \cite{master} although at smaller ranking statistics. For the \gls{snr} output, this change lined up well with the lower limit of the \gls{snr} contained in the training samples. This may be a hint that the network presented in \cite{master} successfully learned the lower bound on the \gls{snr} in the training set.

For high \gls{far}s, both outputs show equivalent sensitivities. At low \gls{far}s, on the other hand, the \gls{snr} output is more sensitive and has non-negligible sensitivities down to a \gls{far} of 10 per month, where it reaches a sensitive radius of \SI{\approx 130}{\mega\parsec}. The sensitivity of the p-score output becomes negligible around a \gls{far} of 20 per month and also reaches a sensitive radius of \SI{\approx 130}{\mega\parsec}. 

In our previous work \cite{master} we observed the opposite behavior with regards to which of the two outputs is more sensitive at low \gls{far}s. We do not know what causes either output to perform better than the other.

We can also observe a change in gradient in the sensitivity curves shown in \autoref{fig:roc}. The locations where the sensitivity starts to suddenly drop steeply line up with the point where the \gls{far} levels off observed in \autoref{fig:false-alarm}. At \gls{far}s below this point the sensitivity becomes negligible quickly.


\subsection{Comparison to PyCBC Live}
We compare our search to PyCBC Live \cite{pycbc_sensitivity_plot}, which is a low-latency analysis and has been used in the second and third observing runs \cite{ligo_pipelines, catalog, gw190412, gw190425}. The green curve in \autoref{fig:roc} is estimated from Figure 1 in \cite{pycbc_sensitivity_plot} on our test set by assuming that all injections with optimal \gls{snr} $>\mathcal{R}$ are found and all others are missed. Here $\mathcal{R}$ is the network \gls{snr} reweighted by signal consistency tests corresponding to a given \gls{far}. At a \gls{far} of $0.5$ per month the PyCBC Live search has a sensitive radius of \SI{\approx 245}{\mega\parsec}. At a comparable \gls{far} of 0.6 per month, our search reaches $1/6$ the sensitive radius. At a \gls{far} of 10 per month, where our search is still marginally sensitive, the radius increases to \SI{\approx 130}{\mega\parsec}, which is still about half the radius of the reference value from the PyCBC Live search. To reach this reference value, we would need to operate at a \gls{far} of $\approx 35,000$ per month.

To compare the computational cost of our search to that of PyCBC Live \cite{pycbc_live}, we analyze the resources both algorithms require to evaluate incoming data in real time. One pass of our network on the correctly whitened,  re-sampled and formatted data takes \SI{206}{\milli\second} on an Intel i7-6600U dual-core, 4 thread laptop processor. Neglecting the cost of pre-processing the raw input data, our search would be able to produce estimates of the \gls{snr} and p-score slightly faster than real time with the above mentioned resources, as each step processes a time span of \SI{250}{\milli\second}.  We can estimate the number of \gls{cpu}-cores PyCBC Live would require to run a search for \gls{bns} signals of the distribution our network was trained on by counting the number of templates the corresponding filter-bank would use. To produce a non-spinning filter-bank with component masses between $1.2$ and $1.6$ solar masses at $3.5$ \gls{pn}-order we use the \verb|pycbc_geom_nonspinbank| program from the PyCBC software library \cite{pycbc}. The minimum match is set to $0.965$. With these settings the bank contains $1960$ templates per detector. The PyCBC Live search is able to evaluate $6300$ templates per core in real time \cite{pycbc_live}. The required $3920$ templates for a 2 detector search could therefore be evaluated on a single core in real time.

At a \gls{far} of 10 per month, our search introduces an average latency of $10.2$ seconds for true positives. This value is calculated by taking the difference between the latest position in any cluster at the given \gls{far} that belongs to a real injection and the reported time of the corresponding event. To ensure that the cluster is complete we add $1$ second on top of that latency and another $0.206$ seconds to ensure the network has finished its calculations. We then average over all clusters evaluated that way. Our search algorithm has not yet been optimized for low-latency analysis and we assume that the latency can be reduced by about an order of magnitude by choosing a different clustering algorithm without a large impact on the sensitivity. The reported latency does not take into account any time lost due to whitening, re-sampling or formatting of the raw input data. PyCBC Live for comparison operates at an average latency of \SI{16}{\second}. This latency can be reduced to \SI{10}{\second} at the cost of doubling the required computational resources \cite{pycbc_live}.

\subsection{Comparison to another machine learning algorithm}
The authors of \cite{bns_network} were the first to search for \gls{bns} signals using a machine learning algorithm.
They used a convolutional network very similar in structure to those found in \cite{original_deep_filtering, hunter} to analyze $10$ seconds of data sampled at \SI{4}{\kilo\hertz}. This setup allowed them to differentiate the three classes ''pure noise'', ''\gls{bbh} signal'' and ''\gls{bns} signal''. They found that their algorithm is able to distinguish the three classes and is as sensitive to \gls{bbh} signals as the previous comparable search algorithms \cite{original_deep_filtering, hunter}. The sensitivity to \gls{bns} signals is below that of \gls{bbh} signals for all signal strengths and false-alarm probabilities tested.

During the preparation of this paper the original pre-print \cite{bns_paper_v1} was re-written and published. In that version the sensitivity to \gls{bns} signals was found to be on par with the sensitivity to \gls{bbh} signals. However, all results were given in terms of peak signal-to-noise ratio (\gls{psnr}) instead of optimal or matched-filter \gls{snr}. The new version removes this hurdle and gives results in terms of optimal \gls{snr}. We therefore comment on both versions.

To convert between \gls{psnr} and matched filter \gls{snr} the authors of \cite{bns_paper_v1} quote a factor of $13$, which was derived on \gls{bbh} data. We calculated this factor on \gls{bns} data and find a conversion of optimal $\text{\gls{snr}}\approx\text{matched-filter \gls{snr}}\approx 41.2\cdot\text{\gls{psnr}}$. Furthermore, they used data from a single detector. Signals detected at \gls{snr} $\rho$ gain on average a factor of $\sqrt{2}$ when a network of $2$ detectors is used. Our results are compared to the findings of \cite{bns_paper_v1} by using the conversion $\text{optimal \gls{snr}}=41.2\cdot\sqrt{2}\cdot\text{\gls{psnr}}$.

The comparison between our work and the results given in \cite{bns_network} still takes the scaling factor of $\sqrt{2}$ into account to compensate for the two-detector setup.
 
\begin{figure}
    \centering
    \includegraphics[width=0.45\textwidth]{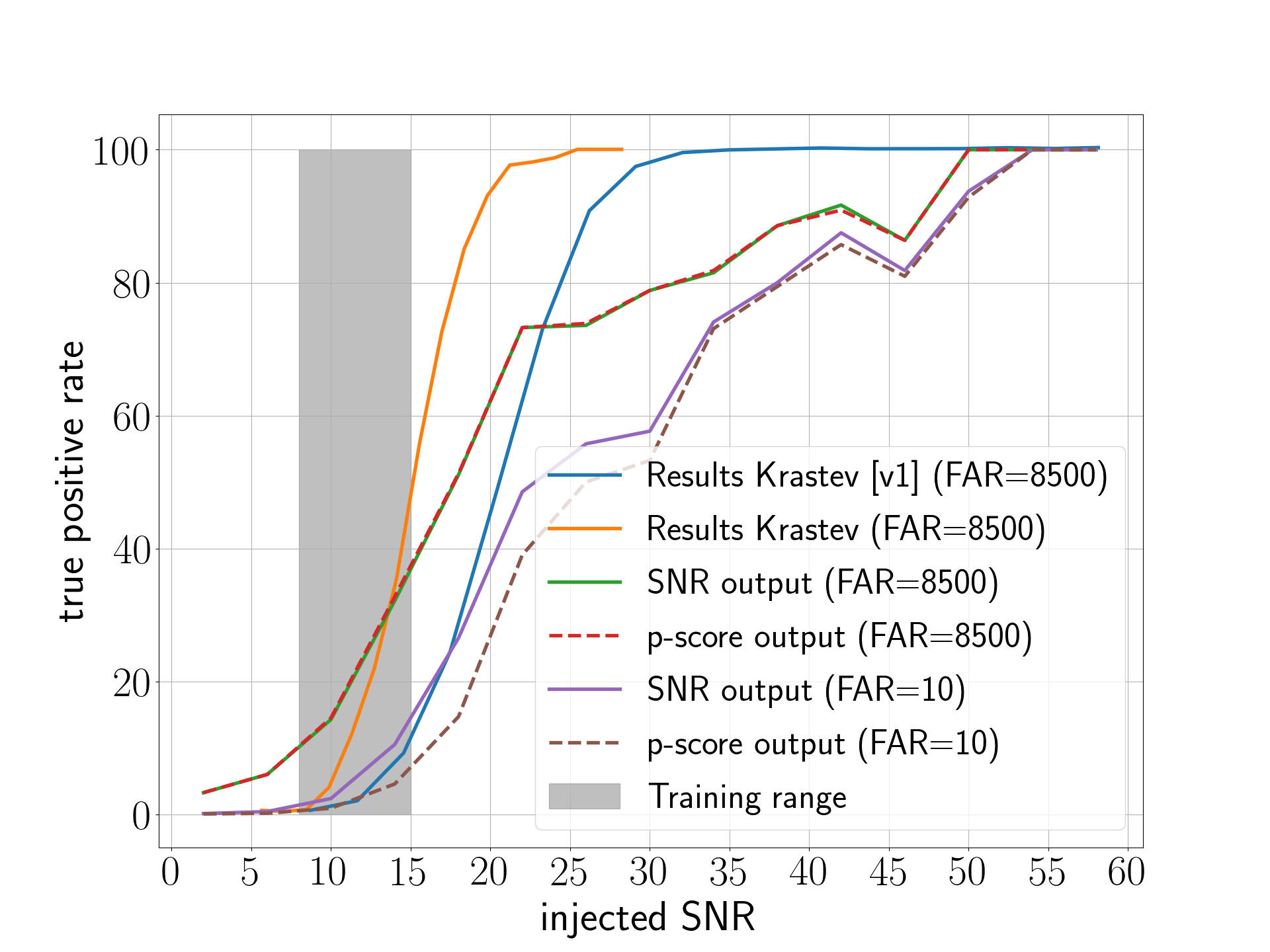}
    \caption{To compare our search to the work of \cite{bns_network} we plot their true positive rate at a fixed \gls{far} of $8500$ per month in yellow and our true positive rate at the same \gls{far} in green and red. On the x-axis we track the injected optimal network \gls{snr}. The blue curve shows the data from \cite{bns_paper_v1}, where the results were given in terms of \gls{psnr}. We use the conversion $\text{\gls{snr}} = 41.2\cdot\sqrt{2}\cdot\text{\gls{psnr}}$. To obtain these curves we bin the injected signals by their optimal injection \gls{snr} and a bin size of 4. For high \gls{snr}s some bins are empty. Empty bins are interpolated linearly from the remaining data. The area marked gray highlights the region covered by the training set. We find that our search performs better for low \gls{snr}s but is less sensitive for strong signals. We also show the true positive rate of our search at a \gls{far} of 10 in purple and brown. Within the training range we find that our search closely matches the true positive rate of \cite{bns_paper_v1} at a higher \gls{far}.}
    \label{fig:comparison_bns_sensitivity}
\end{figure}

\autoref{fig:comparison_bns_sensitivity} compares the true positive rates of the search algorithm presented here to the one found in \cite{bns_network} and \cite{bns_paper_v1} at a fixed \gls{far} of $8500$ per month.\footnote{In \cite{bns_network} no \gls{far} is stated explicitly. All results are given in terms of a false-alarm probability. We estimate the \gls{far} from this probability by dividing with the step-size used to slide the network across long stretches of data. We then re-scale it to false alarms per month. The step-size was estimated to be \SI{0.3}{\s} to match the interval duration within which the peak amplitude for \gls{bns} signals is varied.} We compute it by fixing the detection threshold to the corresponding values of $\text{\gls{snr}}\approx 6.53$ and $\text{p-score}\approx 0.185$. The injections are then binned by their \gls{snr} and for each bin the fraction of detected over total injections is calculated. We find that our search does not generalize well to very strong signals. The loudest missed signal at this \gls{far} was injected with a \gls{snr} of $46.65$ which means that our search only reaches $100\%$ sensitivity above \gls{snr} $46.65$. The search described in \cite{bns_paper_v1} is more sensitive to signals above \gls{snr} $25$ and saturates already around \gls{snr} $35$. The algorithm described in \cite{bns_network} saturates even earlier at \gls{snr} $25$. For current detectors on the other hand, most signals are expected to be measured with low \gls{snr} \cite{signal_distribution}. Within the \gls{snr}-range covered by the training set (marked gray in \autoref{fig:comparison_bns_sensitivity}), our search is almost 10 times as sensitive when compared to \cite{bns_paper_v1} and still about 4 times as sensitive when compared to \cite{bns_network}.

As the authors of \cite{bns_network} successfully applied curriculum learning, we would expect an increase in sensitivity at high \gls{snr}s if the range in our training set were expanded to include high \gls{snr} signals.

The plot also shows the true positive rate of our network at a \gls{far} of 10 per month, where, within the \gls{snr} range of the training set, the \gls{snr} output roughly matches the true positive rate of the algorithm proposed in \cite{bns_paper_v1} at a 85 times higher \gls{far}. The network proposed in \cite{bns_network} is more sensitive to signals with $\text{\gls{snr}}>8$ when compared to our network operating at a \gls{far} of 10 per month. One can also observe that at a \gls{far} of 10 per month the p-score output is significantly worst over the entire range of injected signals.

\begin{figure}
    \centering
    \includegraphics[width=0.45\textwidth]{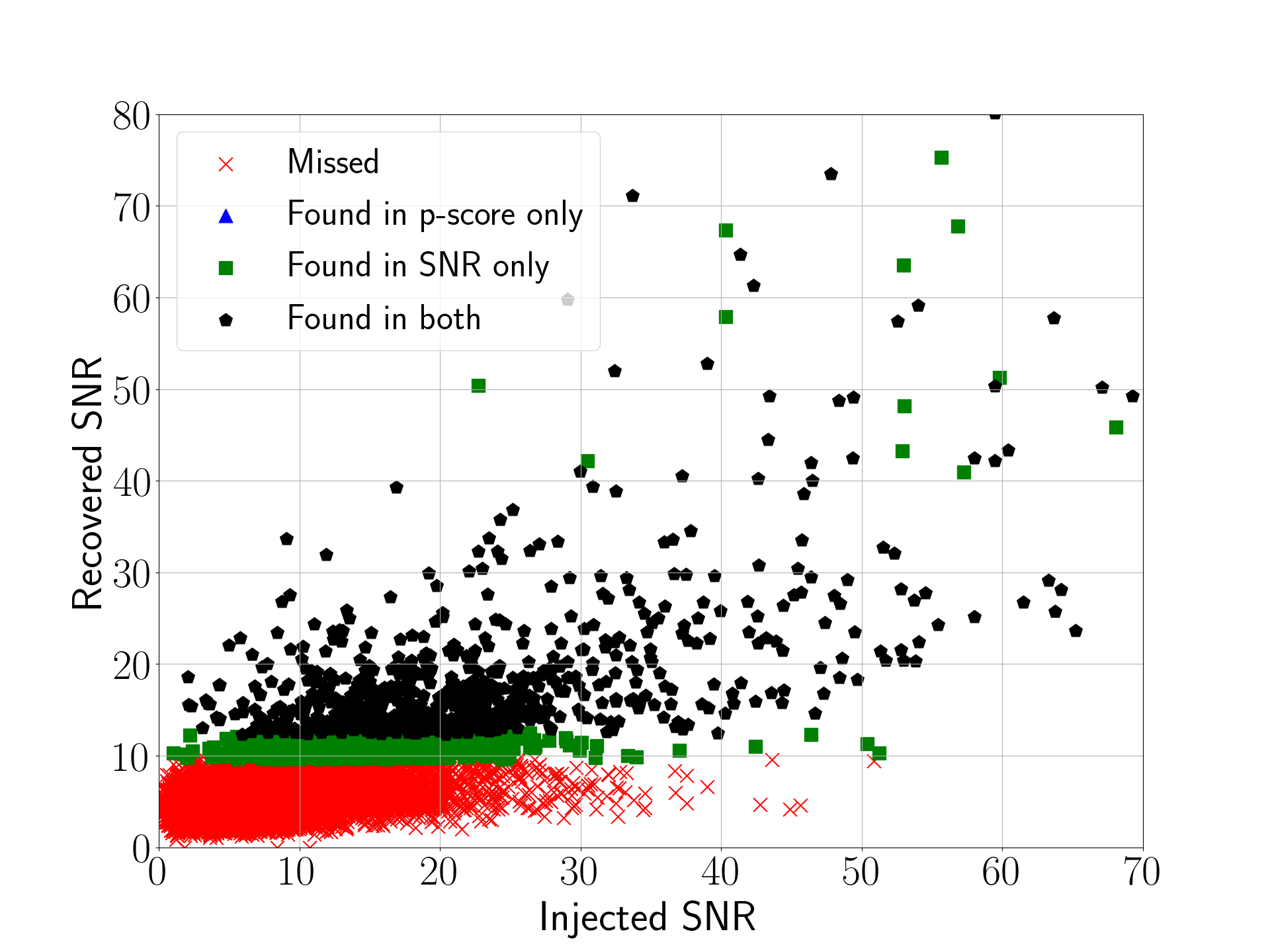}
    \caption{The plot shows the estimated \gls{snr} against the optimal injected \gls{snr} of the test set. There are a few injections with $\text{\gls{snr}}>70$ which are not shown here, but all of them are detected. The red cross corresponds to injections that the search did not recover in either of the two outputs at a fixed \gls{far} of 10 per month. Injections that are found in the p-score output but not in the \gls{snr} output would be shown as blue triangles, but no injections of this type exist. Green squares show injections that are found only in the \gls{snr} output. Black hexagons represent injections that were found in both outputs. A clear vertical separation can be seen in this figure. We suspect that the network learns an estimate of the \gls{snr} internally and only maps the p-score to this internal representation. Otherwise borders would not be so sharp and some blue triangles should be seen in the red area.}
    \label{fig:snr_scatter}
\end{figure}

\autoref{fig:snr_scatter} shows the recovered \gls{snr} against the injected \gls{snr} at a fixed \gls{far} of 10 per month. For any missed injection, we give the value of the estimated \gls{snr} time series that is closest to the injection time. The strongest missed injection at this \gls{far} has a \gls{snr} of $50.83$. We find that the injected \gls{snr} is recovered with a mean squared error of $\approx 181$. Our search is therefore able to distinguish signals from the background but the estimation of the \gls{snr} is uninformative. At this \gls{far}, there are no injections that are only detected in the p-score output. The plot can visually be split into three vertical zones. The lowest zone (red) contains all missed injections. They are recovered with a \gls{snr} below the threshold for the \gls{snr} output. If the p-score output was independent of the \gls{snr} output we would expect to find a few blue triangles in this region. The second zone (green) contains injections that are only recovered in the \gls{snr} output. The clear separation to the third zone (black) indicates that the p-score output operates very similarly to the \gls{snr} output and assigns a value based on the internal \gls{snr} estimate. The louder the injected signals are the more likely the network is to detect it in both outputs.
 
 \subsection{Binary black hole injections}
Realistic continuous data will not only contain signals from \gls{bns} mergers but also prominently signals from \gls{bbh} events. It is therefore interesting to test the response of the \gls{nn} to these kinds of signals.

To do so, we generate a mock data set containing \gls{bbh} injections. We use the process described in \ref{sec:evaluateion_general} but adjust the parameter ranges to represent a distribution for \gls{bbh} signals. The masses are uniformly distributed in the range from 10 to 30 solar masses, the maximal distance is increased to \SI{4000}{\mega\parsec} to adjust for the louder signals and the waveform model is changed to \verb|SEOBNRv4_opt|. As signals from \gls{bbh}s are within the sensitive band of the detectors for a shorter duration, the average signal separation can be reduced to 20 seconds with a variance of $\pm 2$ seconds. The duration of the sections in the beginning and end of each chunk that do not contain any injections is reduced to 16 seconds.

As we only want to make a qualitative statement about the sensitivity of our analysis to \gls{bbh} signals, we generated and evaluated $40960 \text{ seconds} \approx 11$ hours of mock data, containing $2147$ injected signals. The data is processed in the same way as the data containing \gls{bns} injections.

For this test set we find that our network has negligible sensitivity to \gls{bbh} mergers.
The \gls{bbh} waveforms, which are short compared to \gls{bns} signals, are consistently classified as noise.

\section{Conclusions\label{sec:conclusion}}

We presented a new machine learning based search pipeline for \gls{bns} signals. To allow our network to efficiently process up to \SI{32}{\second} of data, we introduced multi-rate sampling; a technique that samples different parts of a \gls{gw} signal at different sample rates.

Our search improves upon the sensitivity of other machine learning based algorithms at low \gls{far}s and for signals with low \gls{snr}. For signals with a \gls{snr} contained within the \gls{snr} boundaries of our training set we find an improvement of $400$\% over previous machine-learning based searches \cite{bns_network}.

We probed, for the first time, the sensitivity of a machine learning based algorithm at \gls{far}s down to $0.5$ per month. This enabled a direct comparison to the template based low-latency search PyCBC Live \cite{pycbc_live}. We found that our machine learning based algorithm is computationally more expensive than using PyCBC Live with a template bank equivalent to our training set. At the same time, the sensitive radius of our search algorithm is lower by a factor of $6$. We therefore conclude that machine learning based search algorithms are not yet sufficiently sensitive or efficient to match the algorithms currently in use.

We do, however, find an improvement in the latency when compared to traditional searches. PyCBC Live introduces on average a latency of $16$s between signal arrival and generating an alert. A very conservative estimate of the latency introduced by our search finds $10.2$s. This value is limited not by the computational cost of applying the network, but by processing the data it outputs. Choosing a different algorithm to do so is straightforward and might improve the latency by roughly an order of magnitude. 
The latency of PyCBC Live can be reduced to a similar duration by increasing the computational cost of the analysis. There are also other search algorithms targeted specifically at low-latency detection of candidate events which are already able to achieve latencies of \SI[parse-numbers=false]{\mathcal{O}(1)}{\second} \cite{spiir_pipeline}. The computational cost of all of these searches scales with the size of the template bank used. \gls{nn}s on the other hand have often proven to adapt well to a large variety of features in the input space. It is therefore not unreasonable to believe that machine learning search algorithms may be able to provide low-latency detections at constant or only slightly increased computational cost when the parameter space is enlarged. We think that this is a strong motivation to further pursue a machine learning based search algorithm.

To help compare different search algorithms we proposed a standardized test procedure that can be applied to neural networks as well. We want to stress the importance of providing \gls{far}s and sensitivities for machine learning based searches which are derived on as realistic a data set as possible.

Future works might try to reduce the complexity of the network proposed here to minimize the computational cost and make machine learning based searches a viable alternative. Reducing the complexity of the network may also help to improve the sensitivity of the search. Previous works \cite{huerta_parameter_estimation} have shown that a network which works well with simulated noise adapts well to real detector noise if re-trained. The algorithm at hand should thus also be extended to be trained and tested on real detector noise. It would further be of interest to test if a computationally less expensive network could be used at a high \gls{far} to be followed up by a matched-filter search with a heavily reduced template bank.

\section{Acknowledgments}
We thank Tobias Blenke, Christoph Drei{\ss}igacker and Tobias Florin for their comments and suggestions. We acknowledge the Max Planck Gesellschaft and the Atlas cluster computing team at Albert-Einstein Institut (AEI) Hannover for support. F.O. was supported by the Max Planck Society's Independent Research Group Programme.

\bibliography{bibliography}

\begin{thebibliography}{66}%
\makeatletter
\providecommand \@ifxundefined [1]{%
 \@ifx{#1\undefined}
}%
\providecommand \@ifnum [1]{%
 \ifnum #1\expandafter \@firstoftwo
 \else \expandafter \@secondoftwo
 \fi
}%
\providecommand \@ifx [1]{%
 \ifx #1\expandafter \@firstoftwo
 \else \expandafter \@secondoftwo
 \fi
}%
\providecommand \natexlab [1]{#1}%
\providecommand \enquote  [1]{``#1''}%
\providecommand \bibnamefont  [1]{#1}%
\providecommand \bibfnamefont [1]{#1}%
\providecommand \citenamefont [1]{#1}%
\providecommand \href@noop [0]{\@secondoftwo}%
\providecommand \href [0]{\begingroup \@sanitize@url \@href}%
\providecommand \@href[1]{\@@startlink{#1}\@@href}%
\providecommand \@@href[1]{\endgroup#1\@@endlink}%
\providecommand \@sanitize@url [0]{\catcode `\\12\catcode `\$12\catcode
  `\&12\catcode `\#12\catcode `\^12\catcode `\_12\catcode `\%12\relax}%
\providecommand \@@startlink[1]{}%
\providecommand \@@endlink[0]{}%
\providecommand \url  [0]{\begingroup\@sanitize@url \@url }%
\providecommand \@url [1]{\endgroup\@href {#1}{\urlprefix }}%
\providecommand \urlprefix  [0]{URL }%
\providecommand \Eprint [0]{\href }%
\providecommand \doibase [0]{http://dx.doi.org/}%
\providecommand \selectlanguage [0]{\@gobble}%
\providecommand \bibinfo  [0]{\@secondoftwo}%
\providecommand \bibfield  [0]{\@secondoftwo}%
\providecommand \translation [1]{[#1]}%
\providecommand \BibitemOpen [0]{}%
\providecommand \bibitemStop [0]{}%
\providecommand \bibitemNoStop [0]{.\EOS\space}%
\providecommand \EOS [0]{\spacefactor3000\relax}%
\providecommand \BibitemShut  [1]{\csname bibitem#1\endcsname}%
\let\auto@bib@innerbib\@empty
\bibitem [{\citenamefont {et~al.}(2016)}]{gw150914}%
  \BibitemOpen
  \bibfield  {author} {\bibinfo {author} {\bibfnamefont {B.~P.~A.}\
  \bibnamefont {et~al.}} (\bibinfo {collaboration} {LIGO Scientific
  Collaboration and Virgo Collaboration}),\ }\href {\doibase
  10.1103/PhysRevLett.116.061102} {\bibfield  {journal} {\bibinfo  {journal}
  {Phys. Rev. Lett.}\ }\textbf {\bibinfo {volume} {116}},\ \bibinfo {pages}
  {061102} (\bibinfo {year} {2016})}\BibitemShut {NoStop}%
\bibitem [{\citenamefont {Abbott}(2019)}]{catalog}%
  \BibitemOpen
  \bibfield  {author} {\bibinfo {author} {\bibfnamefont {B.~P. e.~a.}\
  \bibnamefont {Abbott}} (\bibinfo {collaboration} {LIGO Scientific
  Collaboration and Virgo Collaboration}),\ }\href {\doibase
  10.1103/PhysRevX.9.031040} {\bibfield  {journal} {\bibinfo  {journal} {Phys.
  Rev. X}\ }\textbf {\bibinfo {volume} {9}},\ \bibinfo {pages} {031040}
  (\bibinfo {year} {2019})}\BibitemShut {NoStop}%
\bibitem [{\citenamefont {Venumadhav}\ \emph {et~al.}(2020)\citenamefont
  {Venumadhav}, \citenamefont {Zackay}, \citenamefont {Roulet}, \citenamefont
  {Dai},\ and\ \citenamefont {Zaldarriaga}}]{ias}%
  \BibitemOpen
  \bibfield  {author} {\bibinfo {author} {\bibfnamefont {T.}~\bibnamefont
  {Venumadhav}}, \bibinfo {author} {\bibfnamefont {B.}~\bibnamefont {Zackay}},
  \bibinfo {author} {\bibfnamefont {J.}~\bibnamefont {Roulet}}, \bibinfo
  {author} {\bibfnamefont {L.}~\bibnamefont {Dai}}, \ and\ \bibinfo {author}
  {\bibfnamefont {M.}~\bibnamefont {Zaldarriaga}},\ }\href {\doibase
  10.1103/PhysRevD.101.083030} {\bibfield  {journal} {\bibinfo  {journal}
  {Phys. Rev. D}\ }\textbf {\bibinfo {volume} {101}},\ \bibinfo {pages}
  {083030} (\bibinfo {year} {2020})}\BibitemShut {NoStop}%
\bibitem [{\citenamefont {Nitz}\ \emph
  {et~al.}(2020{\natexlab{a}})\citenamefont {Nitz}, \citenamefont {Dent},
  \citenamefont {Davies}, \citenamefont {Kumar}, \citenamefont {Capano},
  \citenamefont {Harry}, \citenamefont {Mozzon}, \citenamefont {Nuttall},
  \citenamefont {Lundgren},\ and\ \citenamefont {T{\'{a}}pai}}]{2ogc}%
  \BibitemOpen
  \bibfield  {author} {\bibinfo {author} {\bibfnamefont {A.~H.}\ \bibnamefont
  {Nitz}}, \bibinfo {author} {\bibfnamefont {T.}~\bibnamefont {Dent}}, \bibinfo
  {author} {\bibfnamefont {G.~S.}\ \bibnamefont {Davies}}, \bibinfo {author}
  {\bibfnamefont {S.}~\bibnamefont {Kumar}}, \bibinfo {author} {\bibfnamefont
  {C.~D.}\ \bibnamefont {Capano}}, \bibinfo {author} {\bibfnamefont
  {I.}~\bibnamefont {Harry}}, \bibinfo {author} {\bibfnamefont
  {S.}~\bibnamefont {Mozzon}}, \bibinfo {author} {\bibfnamefont
  {L.}~\bibnamefont {Nuttall}}, \bibinfo {author} {\bibfnamefont
  {A.}~\bibnamefont {Lundgren}}, \ and\ \bibinfo {author} {\bibfnamefont
  {M.}~\bibnamefont {T{\'{a}}pai}},\ }\href {\doibase 10.3847/1538-4357/ab733f}
  {\bibfield  {journal} {\bibinfo  {journal} {The Astrophysical Journal}\
  }\textbf {\bibinfo {volume} {891}},\ \bibinfo {pages} {123} (\bibinfo {year}
  {2020}{\natexlab{a}})}\BibitemShut {NoStop}%
\bibitem [{\citenamefont {Nitz}\ \emph {et~al.}(2019)\citenamefont {Nitz},
  \citenamefont {Capano}, \citenamefont {Nielsen}, \citenamefont {Reyes},
  \citenamefont {White}, \citenamefont {Brown},\ and\ \citenamefont
  {Krishnan}}]{Nitz:2018imz}%
  \BibitemOpen
  \bibfield  {author} {\bibinfo {author} {\bibfnamefont {A.~H.}\ \bibnamefont
  {Nitz}}, \bibinfo {author} {\bibfnamefont {C.}~\bibnamefont {Capano}},
  \bibinfo {author} {\bibfnamefont {A.~B.}\ \bibnamefont {Nielsen}}, \bibinfo
  {author} {\bibfnamefont {S.}~\bibnamefont {Reyes}}, \bibinfo {author}
  {\bibfnamefont {R.}~\bibnamefont {White}}, \bibinfo {author} {\bibfnamefont
  {D.~A.}\ \bibnamefont {Brown}}, \ and\ \bibinfo {author} {\bibfnamefont
  {B.}~\bibnamefont {Krishnan}},\ }\href {\doibase 10.3847/1538-4357/ab0108}
  {\bibfield  {journal} {\bibinfo  {journal} {The Astrophysical Journal}\
  }\textbf {\bibinfo {volume} {872}},\ \bibinfo {pages} {195} (\bibinfo {year}
  {2019})}\BibitemShut {NoStop}%
\bibitem [{\citenamefont {Collaboration}\ and\ \citenamefont
  {Collaboration}()}]{o3_alerts}%
  \BibitemOpen
  \bibfield  {author} {\bibinfo {author} {\bibfnamefont {L.~S.}\ \bibnamefont
  {Collaboration}}\ and\ \bibinfo {author} {\bibfnamefont {V.}~\bibnamefont
  {Collaboration}},\ }\href {https://gracedb.ligo.org/superevents/public/O3/}
  {\enquote {\bibinfo {title} {Gracedb — gravitational-wave candidate event
  database},}\ }\BibitemShut {NoStop}%
\bibitem [{\citenamefont {Collaboration}\ and\ \citenamefont {the
  Virgo~Collaboration}(2020)}]{gw190412}%
  \BibitemOpen
  \bibfield  {author} {\bibinfo {author} {\bibfnamefont {T.~L.~S.}\
  \bibnamefont {Collaboration}}\ and\ \bibinfo {author} {\bibnamefont {the
  Virgo~Collaboration}},\ }\href {https://arxiv.org/abs/2004.08342} {\
  (\bibinfo {year} {2020})},\ \Eprint {http://arxiv.org/abs/2004.08342}
  {arXiv:2004.08342 [astro-ph.HE]} \BibitemShut {NoStop}%
\bibitem [{\citenamefont {et~al.}(2020{\natexlab{a}})}]{gw190425}%
  \BibitemOpen
  \bibfield  {author} {\bibinfo {author} {\bibfnamefont {B.~P.~A.}\
  \bibnamefont {et~al.}},\ }\href {\doibase 10.3847/2041-8213/ab75f5}
  {\bibfield  {journal} {\bibinfo  {journal} {The Astrophysical Journal}\
  }\textbf {\bibinfo {volume} {892}},\ \bibinfo {pages} {L3} (\bibinfo {year}
  {2020}{\natexlab{a}})}\BibitemShut {NoStop}%
\bibitem [{\citenamefont {Abbott}\ and\ \citenamefont
  {et~al.}(2020)}]{gw190814}%
  \BibitemOpen
  \bibfield  {author} {\bibinfo {author} {\bibfnamefont {R.}~\bibnamefont
  {Abbott}}\ and\ \bibinfo {author} {\bibnamefont {et~al.}},\ }\href {\doibase
  10.3847/2041-8213/ab960f} {\bibfield  {journal} {\bibinfo  {journal} {The
  Astrophysical Journal}\ }\textbf {\bibinfo {volume} {896}},\ \bibinfo {pages}
  {L44} (\bibinfo {year} {2020})}\BibitemShut {NoStop}%
\bibitem [{\citenamefont {Abbott}(2020)}]{gw190521}%
  \BibitemOpen
  \bibfield  {author} {\bibinfo {author} {\bibfnamefont {R.~e.~a.}\
  \bibnamefont {Abbott}} (\bibinfo {collaboration} {LIGO Scientific
  Collaboration and Virgo Collaboration}),\ }\href {\doibase
  10.1103/PhysRevLett.125.101102} {\bibfield  {journal} {\bibinfo  {journal}
  {Phys. Rev. Lett.}\ }\textbf {\bibinfo {volume} {125}},\ \bibinfo {pages}
  {101102} (\bibinfo {year} {2020})}\BibitemShut {NoStop}%
\bibitem [{\citenamefont {Aso}\ \emph {et~al.}(2013)\citenamefont {Aso},
  \citenamefont {Michimura}, \citenamefont {Somiya}, \citenamefont {Ando},
  \citenamefont {Miyakawa}, \citenamefont {Sekiguchi}, \citenamefont
  {Tatsumi},\ and\ \citenamefont {Yamamoto}}]{kagra}%
  \BibitemOpen
  \bibfield  {author} {\bibinfo {author} {\bibfnamefont {Y.}~\bibnamefont
  {Aso}}, \bibinfo {author} {\bibfnamefont {Y.}~\bibnamefont {Michimura}},
  \bibinfo {author} {\bibfnamefont {K.}~\bibnamefont {Somiya}}, \bibinfo
  {author} {\bibfnamefont {M.}~\bibnamefont {Ando}}, \bibinfo {author}
  {\bibfnamefont {O.}~\bibnamefont {Miyakawa}}, \bibinfo {author}
  {\bibfnamefont {T.}~\bibnamefont {Sekiguchi}}, \bibinfo {author}
  {\bibfnamefont {D.}~\bibnamefont {Tatsumi}}, \ and\ \bibinfo {author}
  {\bibfnamefont {H.}~\bibnamefont {Yamamoto}} (\bibinfo {collaboration} {The
  KAGRA Collaboration}),\ }\href {\doibase 10.1103/PhysRevD.88.043007}
  {\bibfield  {journal} {\bibinfo  {journal} {Phys. Rev. D}\ }\textbf {\bibinfo
  {volume} {88}},\ \bibinfo {pages} {043007} (\bibinfo {year}
  {2013})}\BibitemShut {NoStop}%
\bibitem [{\citenamefont {et~al.}(2020{\natexlab{b}})}]{kagra2}%
  \BibitemOpen
  \bibfield  {author} {\bibinfo {author} {\bibfnamefont {T.~A.}\ \bibnamefont
  {et~al.}},\ }\href {https://arxiv.org/abs/2005.05574} {\enquote {\bibinfo
  {title} {Overview of kagra: Detector design and construction history},}\ }
  (\bibinfo {year} {2020}{\natexlab{b}}),\ \Eprint
  {http://arxiv.org/abs/2005.05574} {arXiv:2005.05574 [physics.ins-det]}
  \BibitemShut {NoStop}%
\bibitem [{\citenamefont {Abbott}(2018)}]{sensitivity_improvments_next_runs}%
  \BibitemOpen
  \bibfield  {author} {\bibinfo {author} {\bibfnamefont {B.~P. e.~a.}\
  \bibnamefont {Abbott}},\ }\href {\doibase 10.1007/s41114-018-0012-9}
  {\bibfield  {journal} {\bibinfo  {journal} {Living Reviews in Relativity}\
  }\textbf {\bibinfo {volume} {21}},\ \bibinfo {pages} {3} (\bibinfo {year}
  {2018})}\BibitemShut {NoStop}%
\bibitem [{\citenamefont {Collaboration}\ and\ \citenamefont
  {Collaboration}(2018)}]{ligo_pipelines}%
  \BibitemOpen
  \bibfield  {author} {\bibinfo {author} {\bibfnamefont {L.~S.}\ \bibnamefont
  {Collaboration}}\ and\ \bibinfo {author} {\bibfnamefont {V.}~\bibnamefont
  {Collaboration}},\ }\href
  {https://emfollow.docs.ligo.org/userguide/analysis/searches.html} {\enquote
  {\bibinfo {title} {Online pipelines},}\ } (\bibinfo {year}
  {2018})\BibitemShut {NoStop}%
\bibitem [{\citenamefont {Maggiore}(2008)}]{gwv1}%
  \BibitemOpen
  \bibfield  {author} {\bibinfo {author} {\bibfnamefont {M.}~\bibnamefont
  {Maggiore}},\ }\href@noop {} {\emph {\bibinfo {title} {Gravitational
  Waves}}}\ (\bibinfo  {publisher} {Oxford University Press},\ \bibinfo {year}
  {2008})\BibitemShut {NoStop}%
\bibitem [{\citenamefont {et~al.}(2019{\natexlab{a}})}]{gstlal_pipeline}%
  \BibitemOpen
  \bibfield  {author} {\bibinfo {author} {\bibfnamefont {S.~S.}\ \bibnamefont
  {et~al.}},\ }\href {https://arxiv.org/abs/1901.08580} {\enquote {\bibinfo
  {title} {The gstlal search analysis methods for compact binary mergers in
  advanced ligo's second and advanced virgo's first observing runs},}\ }
  (\bibinfo {year} {2019}{\natexlab{a}}),\ \Eprint
  {http://arxiv.org/abs/1901.08580} {arXiv:1901.08580} \BibitemShut {NoStop}%
\bibitem [{\citenamefont {Adams}\ \emph {et~al.}(2016)\citenamefont {Adams},
  \citenamefont {Buskulic}, \citenamefont {Germain}, \citenamefont {Guidi},
  \citenamefont {Marion}, \citenamefont {Montani}, \citenamefont {Mours},
  \citenamefont {Piergiovanni},\ and\ \citenamefont {Wang}}]{mbta_pipeline}%
  \BibitemOpen
  \bibfield  {author} {\bibinfo {author} {\bibfnamefont {T.}~\bibnamefont
  {Adams}}, \bibinfo {author} {\bibfnamefont {D.}~\bibnamefont {Buskulic}},
  \bibinfo {author} {\bibfnamefont {V.}~\bibnamefont {Germain}}, \bibinfo
  {author} {\bibfnamefont {G.~M.}\ \bibnamefont {Guidi}}, \bibinfo {author}
  {\bibfnamefont {F.}~\bibnamefont {Marion}}, \bibinfo {author} {\bibfnamefont
  {M.}~\bibnamefont {Montani}}, \bibinfo {author} {\bibfnamefont
  {B.}~\bibnamefont {Mours}}, \bibinfo {author} {\bibfnamefont
  {F.}~\bibnamefont {Piergiovanni}}, \ and\ \bibinfo {author} {\bibfnamefont
  {G.}~\bibnamefont {Wang}},\ }\href {\doibase 10.1088/0264-9381/33/17/175012}
  {\bibfield  {journal} {\bibinfo  {journal} {Classical and Quantum Gravity}\
  }\textbf {\bibinfo {volume} {33}},\ \bibinfo {pages} {175012} (\bibinfo
  {year} {2016})}\BibitemShut {NoStop}%
\bibitem [{\citenamefont {Nitz}\ \emph {et~al.}(2018)\citenamefont {Nitz},
  \citenamefont {Dal~Canton}, \citenamefont {Davis},\ and\ \citenamefont
  {Reyes}}]{pycbc_live}%
  \BibitemOpen
  \bibfield  {author} {\bibinfo {author} {\bibfnamefont {A.~H.}\ \bibnamefont
  {Nitz}}, \bibinfo {author} {\bibfnamefont {T.}~\bibnamefont {Dal~Canton}},
  \bibinfo {author} {\bibfnamefont {D.}~\bibnamefont {Davis}}, \ and\ \bibinfo
  {author} {\bibfnamefont {S.}~\bibnamefont {Reyes}},\ }\href {\doibase
  10.1103/PhysRevD.98.024050} {\bibfield  {journal} {\bibinfo  {journal} {Phys.
  Rev. D}\ }\textbf {\bibinfo {volume} {98}},\ \bibinfo {pages} {024050}
  (\bibinfo {year} {2018})}\BibitemShut {NoStop}%
\bibitem [{\citenamefont {Hooper}\ \emph {et~al.}(2012)\citenamefont {Hooper},
  \citenamefont {Chung}, \citenamefont {Luan}, \citenamefont {Blair},
  \citenamefont {Chen},\ and\ \citenamefont {Wen}}]{spiir_pipeline}%
  \BibitemOpen
  \bibfield  {author} {\bibinfo {author} {\bibfnamefont {S.}~\bibnamefont
  {Hooper}}, \bibinfo {author} {\bibfnamefont {S.~K.}\ \bibnamefont {Chung}},
  \bibinfo {author} {\bibfnamefont {J.}~\bibnamefont {Luan}}, \bibinfo {author}
  {\bibfnamefont {D.}~\bibnamefont {Blair}}, \bibinfo {author} {\bibfnamefont
  {Y.}~\bibnamefont {Chen}}, \ and\ \bibinfo {author} {\bibfnamefont
  {L.}~\bibnamefont {Wen}},\ }\href {\doibase 10.1103/PhysRevD.86.024012}
  {\bibfield  {journal} {\bibinfo  {journal} {Phys. Rev. D}\ }\textbf {\bibinfo
  {volume} {86}},\ \bibinfo {pages} {024012} (\bibinfo {year}
  {2012})}\BibitemShut {NoStop}%
\bibitem [{\citenamefont {Canton}\ and\ \citenamefont
  {Harry}(2017)}]{template_number}%
  \BibitemOpen
  \bibfield  {author} {\bibinfo {author} {\bibfnamefont {T.~D.}\ \bibnamefont
  {Canton}}\ and\ \bibinfo {author} {\bibfnamefont {I.~W.}\ \bibnamefont
  {Harry}},\ }\href {https://arxiv.org/abs/1705.01845} {\enquote {\bibinfo
  {title} {Designing a template bank to observe compact binary coalescences in
  advanced ligo's second observing run},}\ } (\bibinfo {year} {2017}),\ \Eprint
  {http://arxiv.org/abs/1705.01845} {arXiv:1705.01845 [gr-qc]} \BibitemShut
  {NoStop}%
\bibitem [{\citenamefont {Harry}\ \emph {et~al.}(2016)\citenamefont {Harry},
  \citenamefont {Privitera}, \citenamefont {Boh\'e},\ and\ \citenamefont
  {Buonanno}}]{precessionSearch}%
  \BibitemOpen
  \bibfield  {author} {\bibinfo {author} {\bibfnamefont {I.}~\bibnamefont
  {Harry}}, \bibinfo {author} {\bibfnamefont {S.}~\bibnamefont {Privitera}},
  \bibinfo {author} {\bibfnamefont {A.}~\bibnamefont {Boh\'e}}, \ and\ \bibinfo
  {author} {\bibfnamefont {A.}~\bibnamefont {Buonanno}},\ }\href {\doibase
  10.1103/PhysRevD.94.024012} {\bibfield  {journal} {\bibinfo  {journal} {Phys.
  Rev. D}\ }\textbf {\bibinfo {volume} {94}},\ \bibinfo {pages} {024012}
  (\bibinfo {year} {2016})}\BibitemShut {NoStop}%
\bibitem [{\citenamefont {Hannam}\ \emph {et~al.}(2014)\citenamefont {Hannam},
  \citenamefont {Schmidt}, \citenamefont {Boh\'e}, \citenamefont {Haegel},
  \citenamefont {Husa}, \citenamefont {Ohme}, \citenamefont {Pratten},\ and\
  \citenamefont {P\"urrer}}]{firstPrecessionPhenom}%
  \BibitemOpen
  \bibfield  {author} {\bibinfo {author} {\bibfnamefont {M.}~\bibnamefont
  {Hannam}}, \bibinfo {author} {\bibfnamefont {P.}~\bibnamefont {Schmidt}},
  \bibinfo {author} {\bibfnamefont {A.}~\bibnamefont {Boh\'e}}, \bibinfo
  {author} {\bibfnamefont {L.}~\bibnamefont {Haegel}}, \bibinfo {author}
  {\bibfnamefont {S.}~\bibnamefont {Husa}}, \bibinfo {author} {\bibfnamefont
  {F.}~\bibnamefont {Ohme}}, \bibinfo {author} {\bibfnamefont {G.}~\bibnamefont
  {Pratten}}, \ and\ \bibinfo {author} {\bibfnamefont {M.}~\bibnamefont
  {P\"urrer}},\ }\href {\doibase 10.1103/PhysRevLett.113.151101} {\bibfield
  {journal} {\bibinfo  {journal} {Phys. Rev. Lett.}\ }\textbf {\bibinfo
  {volume} {113}},\ \bibinfo {pages} {151101} (\bibinfo {year}
  {2014})}\BibitemShut {NoStop}%
\bibitem [{\citenamefont {Khan}\ \emph {et~al.}(2019)\citenamefont {Khan},
  \citenamefont {Chatziioannou}, \citenamefont {Hannam},\ and\ \citenamefont
  {Ohme}}]{latestPrecessionPhenom}%
  \BibitemOpen
  \bibfield  {author} {\bibinfo {author} {\bibfnamefont {S.}~\bibnamefont
  {Khan}}, \bibinfo {author} {\bibfnamefont {K.}~\bibnamefont {Chatziioannou}},
  \bibinfo {author} {\bibfnamefont {M.}~\bibnamefont {Hannam}}, \ and\ \bibinfo
  {author} {\bibfnamefont {F.}~\bibnamefont {Ohme}},\ }\href {\doibase
  10.1103/PhysRevD.100.024059} {\bibfield  {journal} {\bibinfo  {journal}
  {Phys. Rev. D}\ }\textbf {\bibinfo {volume} {100}},\ \bibinfo {pages}
  {024059} (\bibinfo {year} {2019})}\BibitemShut {NoStop}%
\bibitem [{\citenamefont {Pan}\ \emph {et~al.}(2014)\citenamefont {Pan},
  \citenamefont {Buonanno}, \citenamefont {Taracchini}, \citenamefont {Kidder},
  \citenamefont {Mrou\'e}, \citenamefont {Pfeiffer}, \citenamefont {Scheel},\
  and\ \citenamefont {Szil\'agyi}}]{firstPrecessionEOB}%
  \BibitemOpen
  \bibfield  {author} {\bibinfo {author} {\bibfnamefont {Y.}~\bibnamefont
  {Pan}}, \bibinfo {author} {\bibfnamefont {A.}~\bibnamefont {Buonanno}},
  \bibinfo {author} {\bibfnamefont {A.}~\bibnamefont {Taracchini}}, \bibinfo
  {author} {\bibfnamefont {L.~E.}\ \bibnamefont {Kidder}}, \bibinfo {author}
  {\bibfnamefont {A.~H.}\ \bibnamefont {Mrou\'e}}, \bibinfo {author}
  {\bibfnamefont {H.~P.}\ \bibnamefont {Pfeiffer}}, \bibinfo {author}
  {\bibfnamefont {M.~A.}\ \bibnamefont {Scheel}}, \ and\ \bibinfo {author}
  {\bibfnamefont {B.}~\bibnamefont {Szil\'agyi}},\ }\href {\doibase
  10.1103/PhysRevD.89.084006} {\bibfield  {journal} {\bibinfo  {journal} {Phys.
  Rev. D}\ }\textbf {\bibinfo {volume} {89}},\ \bibinfo {pages} {084006}
  (\bibinfo {year} {2014})}\BibitemShut {NoStop}%
\bibitem [{\citenamefont {Harry}\ \emph {et~al.}(2018)\citenamefont {Harry},
  \citenamefont {Bustillo},\ and\ \citenamefont {Nitz}}]{hmNeglected}%
  \BibitemOpen
  \bibfield  {author} {\bibinfo {author} {\bibfnamefont {I.}~\bibnamefont
  {Harry}}, \bibinfo {author} {\bibfnamefont {J.~C.}\ \bibnamefont {Bustillo}},
  \ and\ \bibinfo {author} {\bibfnamefont {A.}~\bibnamefont {Nitz}},\ }\href
  {\doibase 10.1103/PhysRevD.97.023004} {\bibfield  {journal} {\bibinfo
  {journal} {Phys. Rev. D}\ }\textbf {\bibinfo {volume} {97}},\ \bibinfo
  {pages} {023004} (\bibinfo {year} {2018})}\BibitemShut {NoStop}%
\bibitem [{\citenamefont {London}\ \emph {et~al.}(2018)\citenamefont {London},
  \citenamefont {Khan}, \citenamefont {Fauchon-Jones}, \citenamefont
  {Garc\'{\i}a}, \citenamefont {Hannam}, \citenamefont {Husa}, \citenamefont
  {Jim\'enez-Forteza}, \citenamefont {Kalaghatgi}, \citenamefont {Ohme},\ and\
  \citenamefont {Pannarale}}]{hmModel}%
  \BibitemOpen
  \bibfield  {author} {\bibinfo {author} {\bibfnamefont {L.}~\bibnamefont
  {London}}, \bibinfo {author} {\bibfnamefont {S.}~\bibnamefont {Khan}},
  \bibinfo {author} {\bibfnamefont {E.}~\bibnamefont {Fauchon-Jones}}, \bibinfo
  {author} {\bibfnamefont {C.}~\bibnamefont {Garc\'{\i}a}}, \bibinfo {author}
  {\bibfnamefont {M.}~\bibnamefont {Hannam}}, \bibinfo {author} {\bibfnamefont
  {S.}~\bibnamefont {Husa}}, \bibinfo {author} {\bibfnamefont {X.}~\bibnamefont
  {Jim\'enez-Forteza}}, \bibinfo {author} {\bibfnamefont {C.}~\bibnamefont
  {Kalaghatgi}}, \bibinfo {author} {\bibfnamefont {F.}~\bibnamefont {Ohme}}, \
  and\ \bibinfo {author} {\bibfnamefont {F.}~\bibnamefont {Pannarale}},\ }\href
  {\doibase 10.1103/PhysRevLett.120.161102} {\bibfield  {journal} {\bibinfo
  {journal} {Phys. Rev. Lett.}\ }\textbf {\bibinfo {volume} {120}},\ \bibinfo
  {pages} {161102} (\bibinfo {year} {2018})}\BibitemShut {NoStop}%
\bibitem [{\citenamefont {Khan}\ \emph {et~al.}(2020)\citenamefont {Khan},
  \citenamefont {Ohme}, \citenamefont {Chatziioannou},\ and\ \citenamefont
  {Hannam}}]{precessionHmPhenom}%
  \BibitemOpen
  \bibfield  {author} {\bibinfo {author} {\bibfnamefont {S.}~\bibnamefont
  {Khan}}, \bibinfo {author} {\bibfnamefont {F.}~\bibnamefont {Ohme}}, \bibinfo
  {author} {\bibfnamefont {K.}~\bibnamefont {Chatziioannou}}, \ and\ \bibinfo
  {author} {\bibfnamefont {M.}~\bibnamefont {Hannam}},\ }\href {\doibase
  10.1103/PhysRevD.101.024056} {\bibfield  {journal} {\bibinfo  {journal}
  {Phys. Rev. D}\ }\textbf {\bibinfo {volume} {101}},\ \bibinfo {pages}
  {024056} (\bibinfo {year} {2020})}\BibitemShut {NoStop}%
\bibitem [{\citenamefont {Cotesta}\ \emph {et~al.}(2018)\citenamefont
  {Cotesta}, \citenamefont {Buonanno}, \citenamefont {Boh\'e}, \citenamefont
  {Taracchini}, \citenamefont {Hinder},\ and\ \citenamefont
  {Ossokine}}]{firstHmEOB}%
  \BibitemOpen
  \bibfield  {author} {\bibinfo {author} {\bibfnamefont {R.}~\bibnamefont
  {Cotesta}}, \bibinfo {author} {\bibfnamefont {A.}~\bibnamefont {Buonanno}},
  \bibinfo {author} {\bibfnamefont {A.}~\bibnamefont {Boh\'e}}, \bibinfo
  {author} {\bibfnamefont {A.}~\bibnamefont {Taracchini}}, \bibinfo {author}
  {\bibfnamefont {I.}~\bibnamefont {Hinder}}, \ and\ \bibinfo {author}
  {\bibfnamefont {S.}~\bibnamefont {Ossokine}},\ }\href {\doibase
  10.1103/PhysRevD.98.084028} {\bibfield  {journal} {\bibinfo  {journal} {Phys.
  Rev. D}\ }\textbf {\bibinfo {volume} {98}},\ \bibinfo {pages} {084028}
  (\bibinfo {year} {2018})}\BibitemShut {NoStop}%
\bibitem [{\citenamefont {Nitz}\ \emph
  {et~al.}(2020{\natexlab{b}})\citenamefont {Nitz}, \citenamefont {Lenon},\
  and\ \citenamefont {Brown}}]{eccentricSearch}%
  \BibitemOpen
  \bibfield  {author} {\bibinfo {author} {\bibfnamefont {A.~H.}\ \bibnamefont
  {Nitz}}, \bibinfo {author} {\bibfnamefont {A.}~\bibnamefont {Lenon}}, \ and\
  \bibinfo {author} {\bibfnamefont {D.~A.}\ \bibnamefont {Brown}},\ }\href
  {\doibase 10.3847/1538-4357/ab6611} {\bibfield  {journal} {\bibinfo
  {journal} {The Astrophysical Journal}\ }\textbf {\bibinfo {volume} {890}},\
  \bibinfo {pages} {1} (\bibinfo {year} {2020}{\natexlab{b}})}\BibitemShut
  {NoStop}%
\bibitem [{\citenamefont {Russakovsky}\ \emph {et~al.}(2015)\citenamefont
  {Russakovsky}, \citenamefont {Deng}, \citenamefont {Su}, \citenamefont
  {Krause}, \citenamefont {Satheesh}, \citenamefont {Ma}, \citenamefont
  {Huang}, \citenamefont {Karpathy}, \citenamefont {Khosla}, \citenamefont
  {Bernstein}, \citenamefont {Berg},\ and\ \citenamefont {Fei-Fei}}]{ILSVRC15}%
  \BibitemOpen
  \bibfield  {author} {\bibinfo {author} {\bibfnamefont {O.}~\bibnamefont
  {Russakovsky}}, \bibinfo {author} {\bibfnamefont {J.}~\bibnamefont {Deng}},
  \bibinfo {author} {\bibfnamefont {H.}~\bibnamefont {Su}}, \bibinfo {author}
  {\bibfnamefont {J.}~\bibnamefont {Krause}}, \bibinfo {author} {\bibfnamefont
  {S.}~\bibnamefont {Satheesh}}, \bibinfo {author} {\bibfnamefont
  {S.}~\bibnamefont {Ma}}, \bibinfo {author} {\bibfnamefont {Z.}~\bibnamefont
  {Huang}}, \bibinfo {author} {\bibfnamefont {A.}~\bibnamefont {Karpathy}},
  \bibinfo {author} {\bibfnamefont {A.}~\bibnamefont {Khosla}}, \bibinfo
  {author} {\bibfnamefont {M.}~\bibnamefont {Bernstein}}, \bibinfo {author}
  {\bibfnamefont {A.~C.}\ \bibnamefont {Berg}}, \ and\ \bibinfo {author}
  {\bibfnamefont {L.}~\bibnamefont {Fei-Fei}},\ }\href {\doibase
  10.1007/s11263-015-0816-y} {\bibfield  {journal} {\bibinfo  {journal}
  {International Journal of Computer Vision (IJCV)}\ }\textbf {\bibinfo
  {volume} {115}},\ \bibinfo {pages} {211} (\bibinfo {year}
  {2015})}\BibitemShut {NoStop}%
\bibitem [{\citenamefont {Oord}\ \emph {et~al.}(2016)\citenamefont {Oord},
  \citenamefont {Dieleman}, \citenamefont {Zen}, \citenamefont {Simonyan},
  \citenamefont {Vinyals}, \citenamefont {Graves}, \citenamefont
  {Kalchbrenner}, \citenamefont {Senior},\ and\ \citenamefont
  {Kavukcuoglu}}]{wavenet}%
  \BibitemOpen
  \bibfield  {author} {\bibinfo {author} {\bibfnamefont {A.~v.~d.}\
  \bibnamefont {Oord}}, \bibinfo {author} {\bibfnamefont {S.}~\bibnamefont
  {Dieleman}}, \bibinfo {author} {\bibfnamefont {H.}~\bibnamefont {Zen}},
  \bibinfo {author} {\bibfnamefont {K.}~\bibnamefont {Simonyan}}, \bibinfo
  {author} {\bibfnamefont {O.}~\bibnamefont {Vinyals}}, \bibinfo {author}
  {\bibfnamefont {A.}~\bibnamefont {Graves}}, \bibinfo {author} {\bibfnamefont
  {N.}~\bibnamefont {Kalchbrenner}}, \bibinfo {author} {\bibfnamefont
  {A.}~\bibnamefont {Senior}}, \ and\ \bibinfo {author} {\bibfnamefont
  {K.}~\bibnamefont {Kavukcuoglu}},\ }\href {https://arxiv.org/abs/1609.03499}
  {\bibfield  {journal} {\bibinfo  {journal} {arXiv preprint arXiv:1609.03499}\
  } (\bibinfo {year} {2016})},\ \Eprint {http://arxiv.org/abs/1609.03499}
  {arXiv:1609.03499} \BibitemShut {NoStop}%
\bibitem [{\citenamefont {Silver}\ \emph {et~al.}(2016)\citenamefont {Silver},
  \citenamefont {Huang}, \citenamefont {Maddison}, \citenamefont {Guez},
  \citenamefont {Sifre}, \citenamefont {Van Den~Driessche}, \citenamefont
  {Schrittwieser}, \citenamefont {Antonoglou}, \citenamefont {Panneershelvam},
  \citenamefont {Lanctot} \emph {et~al.}}]{alpha_go}%
  \BibitemOpen
  \bibfield  {author} {\bibinfo {author} {\bibfnamefont {D.}~\bibnamefont
  {Silver}}, \bibinfo {author} {\bibfnamefont {A.}~\bibnamefont {Huang}},
  \bibinfo {author} {\bibfnamefont {C.~J.}\ \bibnamefont {Maddison}}, \bibinfo
  {author} {\bibfnamefont {A.}~\bibnamefont {Guez}}, \bibinfo {author}
  {\bibfnamefont {L.}~\bibnamefont {Sifre}}, \bibinfo {author} {\bibfnamefont
  {G.}~\bibnamefont {Van Den~Driessche}}, \bibinfo {author} {\bibfnamefont
  {J.}~\bibnamefont {Schrittwieser}}, \bibinfo {author} {\bibfnamefont
  {I.}~\bibnamefont {Antonoglou}}, \bibinfo {author} {\bibfnamefont
  {V.}~\bibnamefont {Panneershelvam}}, \bibinfo {author} {\bibfnamefont
  {M.}~\bibnamefont {Lanctot}},  \emph {et~al.},\ }\href {\doibase
  10.1038/nature16961} {\bibfield  {journal} {\bibinfo  {journal} {nature}\
  }\textbf {\bibinfo {volume} {529}},\ \bibinfo {pages} {484} (\bibinfo {year}
  {2016})}\BibitemShut {NoStop}%
\bibitem [{ope(2018)}]{open_ai_five}%
  \BibitemOpen
  \href {https://openai.com/blog/openai-five/} {\enquote {\bibinfo {title}
  {Openai five},}\ } (\bibinfo {year} {2018})\BibitemShut {NoStop}%
\bibitem [{\citenamefont {Bahaadini}\ \emph {et~al.}(2018)\citenamefont
  {Bahaadini}, \citenamefont {Noroozi}, \citenamefont {Rohani}, \citenamefont
  {Coughlin}, \citenamefont {Zevin}, \citenamefont {Smith}, \citenamefont
  {Kalogera},\ and\ \citenamefont {Katsaggelos}}]{gravityspy}%
  \BibitemOpen
  \bibfield  {author} {\bibinfo {author} {\bibfnamefont {S.}~\bibnamefont
  {Bahaadini}}, \bibinfo {author} {\bibfnamefont {V.}~\bibnamefont {Noroozi}},
  \bibinfo {author} {\bibfnamefont {N.}~\bibnamefont {Rohani}}, \bibinfo
  {author} {\bibfnamefont {S.}~\bibnamefont {Coughlin}}, \bibinfo {author}
  {\bibfnamefont {M.}~\bibnamefont {Zevin}}, \bibinfo {author} {\bibfnamefont
  {J.}~\bibnamefont {Smith}}, \bibinfo {author} {\bibfnamefont
  {V.}~\bibnamefont {Kalogera}}, \ and\ \bibinfo {author} {\bibfnamefont
  {A.}~\bibnamefont {Katsaggelos}},\ }\href {\doibase
  10.1016/j.ins.2018.02.068} {\bibfield  {journal} {\bibinfo  {journal}
  {Information Sciences}\ }\textbf {\bibinfo {volume} {444}},\ \bibinfo {pages}
  {172 } (\bibinfo {year} {2018})}\BibitemShut {NoStop}%
\bibitem [{\citenamefont {Dreissigacker}\ \emph {et~al.}(2019)\citenamefont
  {Dreissigacker}, \citenamefont {Sharma}, \citenamefont {Messenger},
  \citenamefont {Zhao},\ and\ \citenamefont {Prix}}]{paper_christoph}%
  \BibitemOpen
  \bibfield  {author} {\bibinfo {author} {\bibfnamefont {C.}~\bibnamefont
  {Dreissigacker}}, \bibinfo {author} {\bibfnamefont {R.}~\bibnamefont
  {Sharma}}, \bibinfo {author} {\bibfnamefont {C.}~\bibnamefont {Messenger}},
  \bibinfo {author} {\bibfnamefont {R.}~\bibnamefont {Zhao}}, \ and\ \bibinfo
  {author} {\bibfnamefont {R.}~\bibnamefont {Prix}},\ }\href {\doibase
  10.1103/PhysRevD.100.044009} {\bibfield  {journal} {\bibinfo  {journal}
  {Phys. Rev. D}\ }\textbf {\bibinfo {volume} {100}},\ \bibinfo {pages}
  {044009} (\bibinfo {year} {2019})}\BibitemShut {NoStop}%
\bibitem [{\citenamefont {Wei}\ and\ \citenamefont
  {Huerta}(2020)}]{dnn_denoising}%
  \BibitemOpen
  \bibfield  {author} {\bibinfo {author} {\bibfnamefont {W.}~\bibnamefont
  {Wei}}\ and\ \bibinfo {author} {\bibfnamefont {E.}~\bibnamefont {Huerta}},\
  }\href {\doibase https://doi.org/10.1016/j.physletb.2019.135081} {\bibfield
  {journal} {\bibinfo  {journal} {Physics Letters B}\ }\textbf {\bibinfo
  {volume} {800}},\ \bibinfo {pages} {135081} (\bibinfo {year}
  {2020})}\BibitemShut {NoStop}%
\bibitem [{\citenamefont {Cuoco}\ \emph {et~al.}(2020)\citenamefont {Cuoco}
  \emph {et~al.}}]{Cuoco:2020ogp}%
  \BibitemOpen
  \bibfield  {author} {\bibinfo {author} {\bibfnamefont {E.}~\bibnamefont
  {Cuoco}} \emph {et~al.},\ }\href@noop {} {\  (\bibinfo {year} {2020})},\
  \Eprint {http://arxiv.org/abs/2005.03745} {arXiv:2005.03745 [astro-ph.HE]}
  \BibitemShut {NoStop}%
\bibitem [{\citenamefont {Green}\ \emph {et~al.}(2020)\citenamefont {Green},
  \citenamefont {Simpson},\ and\ \citenamefont {Gair}}]{Green:2020hst}%
  \BibitemOpen
  \bibfield  {author} {\bibinfo {author} {\bibfnamefont {S.~R.}\ \bibnamefont
  {Green}}, \bibinfo {author} {\bibfnamefont {C.}~\bibnamefont {Simpson}}, \
  and\ \bibinfo {author} {\bibfnamefont {J.}~\bibnamefont {Gair}},\ }\href@noop
  {} {\  (\bibinfo {year} {2020})},\ \Eprint {http://arxiv.org/abs/2002.07656}
  {arXiv:2002.07656 [astro-ph.IM]} \BibitemShut {NoStop}%
\bibitem [{\citenamefont {Gabbard}\ \emph {et~al.}(2019)\citenamefont
  {Gabbard}, \citenamefont {Messenger}, \citenamefont {Heng}, \citenamefont
  {Tonolini},\ and\ \citenamefont {Murray-Smith}}]{Gabbard:2019rde}%
  \BibitemOpen
  \bibfield  {author} {\bibinfo {author} {\bibfnamefont {H.}~\bibnamefont
  {Gabbard}}, \bibinfo {author} {\bibfnamefont {C.}~\bibnamefont {Messenger}},
  \bibinfo {author} {\bibfnamefont {I.~S.}\ \bibnamefont {Heng}}, \bibinfo
  {author} {\bibfnamefont {F.}~\bibnamefont {Tonolini}}, \ and\ \bibinfo
  {author} {\bibfnamefont {R.}~\bibnamefont {Murray-Smith}},\ }\href@noop {} {\
   (\bibinfo {year} {2019})},\ \Eprint {http://arxiv.org/abs/1909.06296}
  {arXiv:1909.06296 [astro-ph.IM]} \BibitemShut {NoStop}%
\bibitem [{\citenamefont {Marulanda}\ \emph {et~al.}(2020)\citenamefont
  {Marulanda}, \citenamefont {Santa},\ and\ \citenamefont
  {Romano}}]{Marulanda:2020nww}%
  \BibitemOpen
  \bibfield  {author} {\bibinfo {author} {\bibfnamefont {J.~P.}\ \bibnamefont
  {Marulanda}}, \bibinfo {author} {\bibfnamefont {C.}~\bibnamefont {Santa}}, \
  and\ \bibinfo {author} {\bibfnamefont {A.~E.}\ \bibnamefont {Romano}},\
  }\href@noop {} {\  (\bibinfo {year} {2020})},\ \Eprint
  {http://arxiv.org/abs/2004.01050} {arXiv:2004.01050 [gr-qc]} \BibitemShut
  {NoStop}%
\bibitem [{\citenamefont {Chan}\ \emph {et~al.}(2020)\citenamefont {Chan},
  \citenamefont {Heng},\ and\ \citenamefont {Messenger}}]{Chan:2019fuz}%
  \BibitemOpen
  \bibfield  {author} {\bibinfo {author} {\bibfnamefont {M.~L.}\ \bibnamefont
  {Chan}}, \bibinfo {author} {\bibfnamefont {I.~S.}\ \bibnamefont {Heng}}, \
  and\ \bibinfo {author} {\bibfnamefont {C.}~\bibnamefont {Messenger}},\ }\href
  {\doibase 10.1103/PhysRevD.102.043022} {\bibfield  {journal} {\bibinfo
  {journal} {Phys. Rev. D}\ }\textbf {\bibinfo {volume} {102}},\ \bibinfo
  {pages} {043022} (\bibinfo {year} {2020})}\BibitemShut {NoStop}%
\bibitem [{\citenamefont {Iess}\ \emph {et~al.}(2020)\citenamefont {Iess},
  \citenamefont {Cuoco}, \citenamefont {Morawski},\ and\ \citenamefont
  {Powell}}]{Iess:2020yqj}%
  \BibitemOpen
  \bibfield  {author} {\bibinfo {author} {\bibfnamefont {A.}~\bibnamefont
  {Iess}}, \bibinfo {author} {\bibfnamefont {E.}~\bibnamefont {Cuoco}},
  \bibinfo {author} {\bibfnamefont {F.}~\bibnamefont {Morawski}}, \ and\
  \bibinfo {author} {\bibfnamefont {J.}~\bibnamefont {Powell}},\ }\href
  {\doibase 10.1088/2632-2153/ab7d31} {\bibfield  {journal} {\bibinfo
  {journal} {Machine Learning: Science and Technology}\ }\textbf {\bibinfo
  {volume} {1}},\ \bibinfo {pages} {025014} (\bibinfo {year}
  {2020})}\BibitemShut {NoStop}%
\bibitem [{\citenamefont {George}\ and\ \citenamefont
  {Huerta}(2018{\natexlab{a}})}]{original_deep_filtering}%
  \BibitemOpen
  \bibfield  {author} {\bibinfo {author} {\bibfnamefont {D.}~\bibnamefont
  {George}}\ and\ \bibinfo {author} {\bibfnamefont {E.~A.}\ \bibnamefont
  {Huerta}},\ }\href {\doibase 10.1103/PhysRevD.97.044039} {\bibfield
  {journal} {\bibinfo  {journal} {Phys. Rev. D}\ }\textbf {\bibinfo {volume}
  {97}},\ \bibinfo {pages} {044039} (\bibinfo {year}
  {2018}{\natexlab{a}})}\BibitemShut {NoStop}%
\bibitem [{\citenamefont {Gabbard}\ \emph {et~al.}(2018)\citenamefont
  {Gabbard}, \citenamefont {Williams}, \citenamefont {Hayes},\ and\
  \citenamefont {Messenger}}]{hunter}%
  \BibitemOpen
  \bibfield  {author} {\bibinfo {author} {\bibfnamefont {H.}~\bibnamefont
  {Gabbard}}, \bibinfo {author} {\bibfnamefont {M.}~\bibnamefont {Williams}},
  \bibinfo {author} {\bibfnamefont {F.}~\bibnamefont {Hayes}}, \ and\ \bibinfo
  {author} {\bibfnamefont {C.}~\bibnamefont {Messenger}},\ }\href {\doibase
  10.1103/PhysRevLett.120.141103} {\bibfield  {journal} {\bibinfo  {journal}
  {Phys. Rev. Lett.}\ }\textbf {\bibinfo {volume} {120}},\ \bibinfo {pages}
  {141103} (\bibinfo {year} {2018})}\BibitemShut {NoStop}%
\bibitem [{\citenamefont {al.}(2016)}]{noise_transients_gw150914}%
  \BibitemOpen
  \bibfield  {author} {\bibinfo {author} {\bibfnamefont {B.~P. A.~E.}\
  \bibnamefont {al.}},\ }\href {\doibase 10.1088/0264-9381/33/13/134001}
  {\bibfield  {journal} {\bibinfo  {journal} {Classical and Quantum Gravity}\
  }\textbf {\bibinfo {volume} {33}},\ \bibinfo {pages} {134001} (\bibinfo
  {year} {2016})}\BibitemShut {NoStop}%
\bibitem [{\citenamefont {Nuttall}(2018)}]{o2_detchar}%
  \BibitemOpen
  \bibfield  {author} {\bibinfo {author} {\bibfnamefont {L.~K.}\ \bibnamefont
  {Nuttall}},\ }\href {\doibase 10.1098/rsta.2017.0286} {\bibfield  {journal}
  {\bibinfo  {journal} {Philosophical Transactions of the Royal Society A:
  Mathematical, Physical and Engineering Sciences}\ }\textbf {\bibinfo {volume}
  {376}},\ \bibinfo {pages} {20170286} (\bibinfo {year} {2018})}\BibitemShut
  {NoStop}%
\bibitem [{\citenamefont {Cabero}\ \emph {et~al.}(2019)\citenamefont {Cabero},
  \citenamefont {Lundgren}, \citenamefont {Nitz}, \citenamefont {Dent},
  \citenamefont {Barker}, \citenamefont {Goetz}, \citenamefont {Kissel},
  \citenamefont {Nuttall}, \citenamefont {Schale}, \citenamefont {Schofield},\
  and\ \citenamefont {Davis}}]{blip_glitches}%
  \BibitemOpen
  \bibfield  {author} {\bibinfo {author} {\bibfnamefont {M.}~\bibnamefont
  {Cabero}}, \bibinfo {author} {\bibfnamefont {A.}~\bibnamefont {Lundgren}},
  \bibinfo {author} {\bibfnamefont {A.~H.}\ \bibnamefont {Nitz}}, \bibinfo
  {author} {\bibfnamefont {T.}~\bibnamefont {Dent}}, \bibinfo {author}
  {\bibfnamefont {D.}~\bibnamefont {Barker}}, \bibinfo {author} {\bibfnamefont
  {E.}~\bibnamefont {Goetz}}, \bibinfo {author} {\bibfnamefont {J.~S.}\
  \bibnamefont {Kissel}}, \bibinfo {author} {\bibfnamefont {L.~K.}\
  \bibnamefont {Nuttall}}, \bibinfo {author} {\bibfnamefont {P.}~\bibnamefont
  {Schale}}, \bibinfo {author} {\bibfnamefont {R.}~\bibnamefont {Schofield}}, \
  and\ \bibinfo {author} {\bibfnamefont {D.}~\bibnamefont {Davis}},\ }\href
  {\doibase 10.1088/1361-6382/ab2e14} {\bibfield  {journal} {\bibinfo
  {journal} {Classical and Quantum Gravity}\ }\textbf {\bibinfo {volume}
  {36}},\ \bibinfo {pages} {155010} (\bibinfo {year} {2019})}\BibitemShut
  {NoStop}%
\bibitem [{\citenamefont {Gebhard}\ \emph {et~al.}(2019)\citenamefont
  {Gebhard}, \citenamefont {Kilbertus}, \citenamefont {Harry},\ and\
  \citenamefont {Sch\"olkopf}}]{cnn_magiacal_bullet}%
  \BibitemOpen
  \bibfield  {author} {\bibinfo {author} {\bibfnamefont {T.~D.}\ \bibnamefont
  {Gebhard}}, \bibinfo {author} {\bibfnamefont {N.}~\bibnamefont {Kilbertus}},
  \bibinfo {author} {\bibfnamefont {I.}~\bibnamefont {Harry}}, \ and\ \bibinfo
  {author} {\bibfnamefont {B.}~\bibnamefont {Sch\"olkopf}},\ }\href {\doibase
  10.1103/PhysRevD.100.063015} {\bibfield  {journal} {\bibinfo  {journal}
  {Phys. Rev. D}\ }\textbf {\bibinfo {volume} {100}},\ \bibinfo {pages}
  {063015} (\bibinfo {year} {2019})}\BibitemShut {NoStop}%
\bibitem [{\citenamefont {Krastev}(2020)}]{bns_network}%
  \BibitemOpen
  \bibfield  {author} {\bibinfo {author} {\bibfnamefont {P.~G.}\ \bibnamefont
  {Krastev}},\ }\href {\doibase https://doi.org/10.1016/j.physletb.2020.135330}
  {\bibfield  {journal} {\bibinfo  {journal} {Physics Letters B}\ }\textbf
  {\bibinfo {volume} {803}},\ \bibinfo {pages} {135330} (\bibinfo {year}
  {2020})}\BibitemShut {NoStop}%
\bibitem [{\citenamefont {Sch\"afer}\ \emph {et~al.}(2020)\citenamefont
  {Sch\"afer}, \citenamefont {Ohme},\ and\ \citenamefont
  {Nitz}}]{bns-ml-release}%
  \BibitemOpen
  \bibfield  {author} {\bibinfo {author} {\bibfnamefont {M.~B.}\ \bibnamefont
  {Sch\"afer}}, \bibinfo {author} {\bibfnamefont {F.}~\bibnamefont {Ohme}}, \
  and\ \bibinfo {author} {\bibfnamefont {A.~H.}\ \bibnamefont {Nitz}},\
  }\href@noop {} {\enquote {\bibinfo {title} {{Data Release: Detection of
  gravitational-wave signals from binary neutron star mergers using machine
  learning}},}\ }\bibinfo {howpublished}
  {\url{https://github.com/gwastro/bns-machine-learning-search}} (\bibinfo
  {year} {2020})\BibitemShut {NoStop}%
\bibitem [{\citenamefont {Usman}\ \emph {et~al.}(2016)\citenamefont {Usman},
  \citenamefont {Nitz}, \citenamefont {Harry}, \citenamefont {Biwer},
  \citenamefont {Brown}, \citenamefont {Cabero}, \citenamefont {Capano},
  \citenamefont {Canton}, \citenamefont {Dent}, \citenamefont {Fairhurst},
  \citenamefont {Kehl}, \citenamefont {Keppel}, \citenamefont {Krishnan},
  \citenamefont {Lenon}, \citenamefont {Lundgren}, \citenamefont {Nielsen},
  \citenamefont {Pekowsky}, \citenamefont {Pfeiffer}, \citenamefont {Saulson},
  \citenamefont {West},\ and\ \citenamefont {Willis}}]{pycbc_search}%
  \BibitemOpen
  \bibfield  {author} {\bibinfo {author} {\bibfnamefont {S.~A.}\ \bibnamefont
  {Usman}}, \bibinfo {author} {\bibfnamefont {A.~H.}\ \bibnamefont {Nitz}},
  \bibinfo {author} {\bibfnamefont {I.~W.}\ \bibnamefont {Harry}}, \bibinfo
  {author} {\bibfnamefont {C.~M.}\ \bibnamefont {Biwer}}, \bibinfo {author}
  {\bibfnamefont {D.~A.}\ \bibnamefont {Brown}}, \bibinfo {author}
  {\bibfnamefont {M.}~\bibnamefont {Cabero}}, \bibinfo {author} {\bibfnamefont
  {C.~D.}\ \bibnamefont {Capano}}, \bibinfo {author} {\bibfnamefont {T.~D.}\
  \bibnamefont {Canton}}, \bibinfo {author} {\bibfnamefont {T.}~\bibnamefont
  {Dent}}, \bibinfo {author} {\bibfnamefont {S.}~\bibnamefont {Fairhurst}},
  \bibinfo {author} {\bibfnamefont {M.~S.}\ \bibnamefont {Kehl}}, \bibinfo
  {author} {\bibfnamefont {D.}~\bibnamefont {Keppel}}, \bibinfo {author}
  {\bibfnamefont {B.}~\bibnamefont {Krishnan}}, \bibinfo {author}
  {\bibfnamefont {A.}~\bibnamefont {Lenon}}, \bibinfo {author} {\bibfnamefont
  {A.}~\bibnamefont {Lundgren}}, \bibinfo {author} {\bibfnamefont {A.~B.}\
  \bibnamefont {Nielsen}}, \bibinfo {author} {\bibfnamefont {L.~P.}\
  \bibnamefont {Pekowsky}}, \bibinfo {author} {\bibfnamefont {H.~P.}\
  \bibnamefont {Pfeiffer}}, \bibinfo {author} {\bibfnamefont {P.~R.}\
  \bibnamefont {Saulson}}, \bibinfo {author} {\bibfnamefont {M.}~\bibnamefont
  {West}}, \ and\ \bibinfo {author} {\bibfnamefont {J.~L.}\ \bibnamefont
  {Willis}},\ }\href {\doibase 10.1088/0264-9381/33/21/215004} {\bibfield
  {journal} {\bibinfo  {journal} {Classical and Quantum Gravity}\ }\textbf
  {\bibinfo {volume} {33}},\ \bibinfo {pages} {215004} (\bibinfo {year}
  {2016})}\BibitemShut {NoStop}%
\bibitem [{\citenamefont {et~al.}(2019{\natexlab{b}})}]{pycbc}%
  \BibitemOpen
  \bibfield  {author} {\bibinfo {author} {\bibfnamefont {A.~N.}\ \bibnamefont
  {et~al.}},\ }\href {\doibase 10.5281/zenodo.2581446} {\enquote {\bibinfo
  {title} {gwastro/pycbc: Pycbc release v1.13.5},}\ } (\bibinfo {year}
  {2019}{\natexlab{b}})\BibitemShut {NoStop}%
\bibitem [{\citenamefont {Droz}\ \emph {et~al.}(1999)\citenamefont {Droz},
  \citenamefont {Knapp}, \citenamefont {Poisson},\ and\ \citenamefont
  {Owen}}]{taylorf2_1}%
  \BibitemOpen
  \bibfield  {author} {\bibinfo {author} {\bibfnamefont {S.}~\bibnamefont
  {Droz}}, \bibinfo {author} {\bibfnamefont {D.~J.}\ \bibnamefont {Knapp}},
  \bibinfo {author} {\bibfnamefont {E.}~\bibnamefont {Poisson}}, \ and\
  \bibinfo {author} {\bibfnamefont {B.~J.}\ \bibnamefont {Owen}},\ }\href
  {\doibase 10.1103/PhysRevD.59.124016} {\bibfield  {journal} {\bibinfo
  {journal} {Phys. Rev. D}\ }\textbf {\bibinfo {volume} {59}},\ \bibinfo
  {pages} {124016} (\bibinfo {year} {1999})}\BibitemShut {NoStop}%
\bibitem [{\citenamefont {Blanchet}(2002)}]{taylorf2_2}%
  \BibitemOpen
  \bibfield  {author} {\bibinfo {author} {\bibfnamefont {L.}~\bibnamefont
  {Blanchet}},\ }\href {\doibase 10.12942/lrr-2002-3} {\bibfield  {journal}
  {\bibinfo  {journal} {Living Reviews in Relativity}\ }\textbf {\bibinfo
  {volume} {5}},\ \bibinfo {pages} {3} (\bibinfo {year} {2002})}\BibitemShut
  {NoStop}%
\bibitem [{\citenamefont {Faye}\ \emph {et~al.}(2012)\citenamefont {Faye},
  \citenamefont {Marsat}, \citenamefont {Blanchet},\ and\ \citenamefont
  {Iyer}}]{taylorf2_3}%
  \BibitemOpen
  \bibfield  {author} {\bibinfo {author} {\bibfnamefont {G.}~\bibnamefont
  {Faye}}, \bibinfo {author} {\bibfnamefont {S.}~\bibnamefont {Marsat}},
  \bibinfo {author} {\bibfnamefont {L.}~\bibnamefont {Blanchet}}, \ and\
  \bibinfo {author} {\bibfnamefont {B.~R.}\ \bibnamefont {Iyer}},\ }\href
  {\doibase 10.1088/0264-9381/29/17/175004} {\bibfield  {journal} {\bibinfo
  {journal} {Classical and Quantum Gravity}\ }\textbf {\bibinfo {volume}
  {29}},\ \bibinfo {pages} {175004} (\bibinfo {year} {2012})}\BibitemShut
  {NoStop}%
\bibitem [{\citenamefont {Collaboration}(2018)}]{lalsuite}%
  \BibitemOpen
  \bibfield  {author} {\bibinfo {author} {\bibfnamefont {L.~S.}\ \bibnamefont
  {Collaboration}},\ }\href {\doibase 10.7935/GT1W-FZ16} {\enquote {\bibinfo
  {title} {{LIGO} {A}lgorithm {L}ibrary - {LALS}uite},}\ }\bibinfo
  {howpublished} {free software (GPL)} (\bibinfo {year} {2018})\BibitemShut
  {NoStop}%
\bibitem [{\citenamefont {George}\ and\ \citenamefont
  {Huerta}(2018{\natexlab{b}})}]{huerta_parameter_estimation}%
  \BibitemOpen
  \bibfield  {author} {\bibinfo {author} {\bibfnamefont {D.}~\bibnamefont
  {George}}\ and\ \bibinfo {author} {\bibfnamefont {E.}~\bibnamefont
  {Huerta}},\ }\href {\doibase 10.1016/j.physletb.2017.12.053} {\bibfield
  {journal} {\bibinfo  {journal} {Physics Letters B}\ }\textbf {\bibinfo
  {volume} {778}},\ \bibinfo {pages} {64 } (\bibinfo {year}
  {2018}{\natexlab{b}})}\BibitemShut {NoStop}%
\bibitem [{\citenamefont {Sch\"afer}(2019)}]{master}%
  \BibitemOpen
  \bibfield  {author} {\bibinfo {author} {\bibfnamefont {M.~B.}\ \bibnamefont
  {Sch\"afer}},\ }\href {\doibase 10.15488/7467} {\enquote {\bibinfo {title}
  {Analysis of gravitational-wave signals from binary neutron star mergers
  using machine learning},}\ } (\bibinfo {year} {2019})\BibitemShut {NoStop}%
\bibitem [{\citenamefont {Schutz}(2011)}]{signal_distribution}%
  \BibitemOpen
  \bibfield  {author} {\bibinfo {author} {\bibfnamefont {B.~F.}\ \bibnamefont
  {Schutz}},\ }\href {\doibase 10.1088/0264-9381/28/12/125023} {\bibfield
  {journal} {\bibinfo  {journal} {Classical and Quantum Gravity}\ }\textbf
  {\bibinfo {volume} {28}},\ \bibinfo {pages} {125023} (\bibinfo {year}
  {2011})}\BibitemShut {NoStop}%
\bibitem [{\citenamefont {{Szegedy}}\ \emph {et~al.}(2015)\citenamefont
  {{Szegedy}}, \citenamefont {{Wei Liu}}, \citenamefont {{Yangqing Jia}},
  \citenamefont {{Sermanet}}, \citenamefont {{Reed}}, \citenamefont
  {{Anguelov}}, \citenamefont {{Erhan}}, \citenamefont {{Vanhoucke}},\ and\
  \citenamefont {{Rabinovich}}}]{inception_module}%
  \BibitemOpen
  \bibfield  {author} {\bibinfo {author} {\bibfnamefont {C.}~\bibnamefont
  {{Szegedy}}}, \bibinfo {author} {\bibnamefont {{Wei Liu}}}, \bibinfo {author}
  {\bibnamefont {{Yangqing Jia}}}, \bibinfo {author} {\bibfnamefont
  {P.}~\bibnamefont {{Sermanet}}}, \bibinfo {author} {\bibfnamefont
  {S.}~\bibnamefont {{Reed}}}, \bibinfo {author} {\bibfnamefont
  {D.}~\bibnamefont {{Anguelov}}}, \bibinfo {author} {\bibfnamefont
  {D.}~\bibnamefont {{Erhan}}}, \bibinfo {author} {\bibfnamefont
  {V.}~\bibnamefont {{Vanhoucke}}}, \ and\ \bibinfo {author} {\bibfnamefont
  {A.}~\bibnamefont {{Rabinovich}}},\ }in\ \href {\doibase
  /10.1109/CVPR.2015.7298594} {\emph {\bibinfo {booktitle} {2015 IEEE
  Conference on Computer Vision and Pattern Recognition (CVPR)}}}\ (\bibinfo
  {year} {2015})\ pp.\ \bibinfo {pages} {1--9}\BibitemShut {NoStop}%
\bibitem [{\citenamefont {Bai}\ \emph {et~al.}(2018)\citenamefont {Bai},
  \citenamefont {Kolter},\ and\ \citenamefont {Koltun}}]{tcn_paper}%
  \BibitemOpen
  \bibfield  {author} {\bibinfo {author} {\bibfnamefont {S.}~\bibnamefont
  {Bai}}, \bibinfo {author} {\bibfnamefont {J.~Z.}\ \bibnamefont {Kolter}}, \
  and\ \bibinfo {author} {\bibfnamefont {V.}~\bibnamefont {Koltun}},\ }\href
  {http://arxiv.org/abs/1803.01271} {\bibfield  {journal} {\bibinfo  {journal}
  {CoRR}\ }\textbf {\bibinfo {volume} {abs/1803.01271}} (\bibinfo {year}
  {2018})},\ \Eprint {http://arxiv.org/abs/1803.01271} {arXiv:1803.01271}
  \BibitemShut {NoStop}%
\bibitem [{\citenamefont {Schmitt}\ \emph {et~al.}(2019)\citenamefont
  {Schmitt}, \citenamefont {Fu}, \citenamefont {Fan},\ and\ \citenamefont
  {Luo}}]{tcn_idea}%
  \BibitemOpen
  \bibfield  {author} {\bibinfo {author} {\bibfnamefont {A.}~\bibnamefont
  {Schmitt}}, \bibinfo {author} {\bibfnamefont {K.}~\bibnamefont {Fu}},
  \bibinfo {author} {\bibfnamefont {S.}~\bibnamefont {Fan}}, \ and\ \bibinfo
  {author} {\bibfnamefont {Y.}~\bibnamefont {Luo}},\ }in\ \href {\doibase
  10.1145/3339363.3339377} {\emph {\bibinfo {booktitle} {Proceedings of the 2Nd
  International Conference on Computer Science and Software Engineering}}},\
  \bibinfo {series and number} {CSSE 2019}\ (\bibinfo  {publisher} {ACM},\
  \bibinfo {address} {New York, NY, USA},\ \bibinfo {year} {2019})\ pp.\
  \bibinfo {pages} {73--78}\BibitemShut {NoStop}%
\bibitem [{\citenamefont {Lin}\ \emph {et~al.}(2013)\citenamefont {Lin},
  \citenamefont {Chen},\ and\ \citenamefont {Yan}}]{dim_red_invention}%
  \BibitemOpen
  \bibfield  {author} {\bibinfo {author} {\bibfnamefont {M.}~\bibnamefont
  {Lin}}, \bibinfo {author} {\bibfnamefont {Q.}~\bibnamefont {Chen}}, \ and\
  \bibinfo {author} {\bibfnamefont {S.}~\bibnamefont {Yan}},\ }\href
  {https://arxiv.org/abs/1312.4400} {\bibfield  {journal} {\bibinfo  {journal}
  {arXiv preprint arXiv:1312.4400}\ } (\bibinfo {year} {2013})},\ \Eprint
  {http://arxiv.org/abs/1312.4400} {arXiv:1312.4400 [cs]} \BibitemShut
  {NoStop}%
\bibitem [{\citenamefont {Chollet}\ \emph {et~al.}(2019)\citenamefont {Chollet}
  \emph {et~al.}}]{keras}%
  \BibitemOpen
  \bibfield  {author} {\bibinfo {author} {\bibfnamefont {F.}~\bibnamefont
  {Chollet}} \emph {et~al.},\ }\href@noop {} {\enquote {\bibinfo {title}
  {Keras},}\ }\bibinfo {howpublished} {\url{https://keras.io}} (\bibinfo {year}
  {2019})\BibitemShut {NoStop}%
\bibitem [{\citenamefont {Nitz}\ \emph {et~al.}(2017)\citenamefont {Nitz},
  \citenamefont {Dent}, \citenamefont {Canton}, \citenamefont {Fairhurst},\
  and\ \citenamefont {Brown}}]{pycbc_sensitivity_plot}%
  \BibitemOpen
  \bibfield  {author} {\bibinfo {author} {\bibfnamefont {A.~H.}\ \bibnamefont
  {Nitz}}, \bibinfo {author} {\bibfnamefont {T.}~\bibnamefont {Dent}}, \bibinfo
  {author} {\bibfnamefont {T.~D.}\ \bibnamefont {Canton}}, \bibinfo {author}
  {\bibfnamefont {S.}~\bibnamefont {Fairhurst}}, \ and\ \bibinfo {author}
  {\bibfnamefont {D.~A.}\ \bibnamefont {Brown}},\ }\href {\doibase
  10.3847/1538-4357/aa8f50} {\bibfield  {journal} {\bibinfo  {journal} {The
  Astrophysical Journal}\ }\textbf {\bibinfo {volume} {849}},\ \bibinfo {pages}
  {118} (\bibinfo {year} {2017})}\BibitemShut {NoStop}%
\bibitem [{\citenamefont {Krastev}(2019)}]{bns_paper_v1}%
  \BibitemOpen
  \bibfield  {author} {\bibinfo {author} {\bibfnamefont {P.~G.}\ \bibnamefont
  {Krastev}},\ }\href {https://arxiv.org/abs/1908.03151v1} {\  (\bibinfo {year}
  {2019})},\ \Eprint {http://arxiv.org/abs/1908.03151v1} {arXiv:1908.03151v1}
  \BibitemShut {NoStop}%
\end{thebibliography}%

\end{document}